\documentclass[11pt,twoside]{article}
\usepackage[utf8]{inputenc}
\usepackage{times,fancyhdr}
\usepackage[dvips]{graphicx}
\usepackage{latexsym}
\usepackage[affil-it]{authblk}
\usepackage{mathptm}
\usepackage{amsmath}
\usepackage{amssymb}
\usepackage{color}
\usepackage{setspace}
\usepackage{hyperref}
\usepackage{floatrow}
\usepackage{longtable}

\usepackage[top=4cm, bottom=4cm, left=4cm, right=4cm]{geometry}

\def\beq{\begin{equation}}
\def\eeq{\end{equation}}
\def\beqn{\begin{eqnarray}}
\def\eeqn{\end{eqnarray}}

\newcommand{\be}{\begin{equation}}
\newcommand{\ee}{\end{equation}}
\newcommand{\bea}{\begin{eqnarray}}
\newcommand{\eea}{\end{eqnarray}}

\begin{document}

\title{Strong lensing with superfluid dark matter}
\author{Sabine Hossenfelder, Tobias Mistele}
\affil{\small Frankfurt Institute for Advanced Studies\\
Ruth-Moufang-Str. 1,
D-60438 Frankfurt am Main, Germany
}
\date{}
\maketitle

\begin{abstract}

In superfluid dark matter the exchange of phonons can create an additional force that
has an effect similar to Modified Newtonian Dynamics ({\sc MOND}).
To test whether this hypothesis is compatible with observation, we study a set of strong gravitational lenses from the {\sc SLACS} survey
and check whether the measurements can be explained by a superfluid
in the central region of galaxies. 

Concretely, we try to simultaneously fit each lens's
Einstein radius and velocity dispersion with a spherically symmetric density profile of a
fluid that has both a normal and a superfluid component. We demonstrate 
that we can successfully fit all galaxies
except one, and that the fits have reasonable stellar 
mass-to-light-ratios. We conclude that strong gravitational
lensing does not pose a challenge for the idea that superfluid
dark matter mimics modified gravity.

\end{abstract}

\section{Introduction}

Recently, a new theory of superfluid dark matter ({\sc SFDM}) has been proposed as a possible solution of the missing-mass problem
on both galactic and cosmological scales \cite{Berezhiani2015, Berezhiani2017}. While the idea that dark matter
made of light particles might condense to a superfluid has been considered before on general grounds 
\cite{Sikivie:2009qn,Noumi:2013zga,Davidson:2013aba,deVega:2014wya,Davidson:2014hfa,Guth:2014hsa,Aguirre:2015mva,Dev:2016hxv,Eby:2017azn,Sarkar:2017aje}, the type
of superfluid proposed in \cite{Berezhiani2015, Berezhiani2017} has an important additional feature: 
The phonons of the superfluid give rise to a new force that resembles the force that was postulated decades ago in 
Modified Newtonal Dynamics ({\sc MOND}) \cite{Milgrom:1983zz,Milgrom:1983ca,Bekenstein:1984tv}.

This type of theory hence occupies an awkward middle position between modified gravity and particle
dark matter. The reason for this hybrid behavior is that on purely mathematical grounds the equations of the superfluid look
very similar to those that were previously discussed as modified gravity. Indeed, it was shown in \cite{Hossenfelder:2017eoh}
that a holography-based ansatz to emergent gravity proposed in \cite{Verlinde:2016toy} results in a generally covariant theory
very similar to the {\sc SFDM} considered in \cite{Berezhiani2015, Berezhiani2017}. The major difference between
the two cases is that the proposal in \cite{Verlinde:2016toy,Hossenfelder:2017eoh} is based on a vector field,
whereas \cite{Berezhiani2015, Berezhiani2017} use a scalar field. For the purposes of the present study,
we use the scalar field. 

In our perspective, it is therefore justified to also  refer to the type of {\sc SFDM} considered here as modified
gravity because -- if this theory is correct -- then the force acting on stars that orbit galactic centers is
not merely due to the gravitational pull of the additional field. The superfluid has a mass and hence excerts
a gravitational pull, but what gives rise to the {\sc MOND}-like properties is the additional phonon force. 
Gravity, hence, is indeed modified.

It is
worthwhile to point out that in the not-superfluid phase the additional force does not exist, so that the
 field acts like normal cold dark matter whenever the temperature is too high or the
potential is not deep enough to induce condensation to a superfluid. While it is not presently known how well 
this agrees with the data, this means there is no reason a priori to worry about the compatibility of the
model considered here with other evidence usually attributed to dark matter, such as the
height of the third acoustic peak of the cosmic microwave background or the dynamics of galactic clusters,
because in neither case the conditions for condensation are fulfilled. 

There is however one observation that springs to mind immediately which might
pose a challenge to {\sc SFDM}, which is gravitational lensing. The reason is the following.

We know
from the detection of a gravitational wave event with optical counterpart \cite{GBM:2017lvd} that gravitational
waves travel at the same speed as photons to high precision. This means that the
phonon force of the superfluid has a negligible, if any, effect on photons \cite{Boran:2017rdn, Sanders:2018jiv}. This is not theoretically
difficult to achieve; it simply means that the respective coupling term does not exist (or is highly
suppressed).
An easy way to achieve this would be, for example, to have a coupling
that depends on the particle's rest mass, in which case photons would remain entirely unaffected
by the phonon force. A similar result could be obtained by postulating that the coupling be conformal.
However, if the total force acting on photons is not the same
as that acting on baryons, this means that the apparent dark matter inferred from the
motion of baryons will generically not match the dark matter inferred from gravitational
lensing. 

The purpose of this present work is to test whether this mismatch can be
detected by use of strong gravitational lensing. The answer, roughly speaking, is ``No'': Strong
gravitational lensing is not a promising way to probe whether {\sc SFDM} mimics
a {\sc MOND}-like force.  The
rest of the paper will explain why that is so and what restrictions apply to this rough 
answer. 

\section{The Model}
\label{model}

We here use the model from \cite{Berezhiani2017} in which the fluid is described by
a massive, real, scalar field $\phi$.  At zero temperature, the Lagrangian has
the form
\beqn
{\cal L}_\phi = 2 \Lambda \frac{(2m)^{3/2}}{3} \chi \sqrt{|\chi|} ~, \label{Lphi}
\eeqn
where $\chi$ is the kinetic term of the scalar field, $m$ is a constant of dimension mass, and $\Lambda$ quantifies
the strength of the self-interaction. In the Newtonian limit the kinetic term can be approximated by
\beqn
\chi \approx \mu - m \Phi + \dot \phi - \frac{1}{2m }(\vec \nabla \phi)^2 ~,
\eeqn
where $\Phi$ is the Newtonian gravitational potential and $\mu$ can be interpreted as the chemical potential.
A dot, as usual, denotes a derivative with respect to time; the gradient contains spatial derivatives only.

In the same limit, the new field $\phi$ couples to the energy-density of baryons $\rho_{\rm b}$ through the interaction term
\beqn
{\cal L}_{\rm int} = \alpha \frac{\Lambda}{m_{\rm pl}} \phi \rho_{\rm b}~, \label{Lint}
\eeqn
where $m_{\rm pl}$ is the Planck mass (and only there for dimensional reasons), and $\alpha$ is a dimensionless
constant that quantifies the strength of the interaction. It is the combination of the power $3/2$ in
the kinetic term (\ref{Lphi}) combined with the peculiar coupling (\ref{Lint}) that gives rise to the {\sc MOND}-like
behavior. The same features can be found in the vector-based model considered in \cite{Hossenfelder:2017eoh}.

If one now derives the equations of motion for the field $\phi$ from the Lagrangian, the
equations will depend on the
Newtonian potential $\Phi$. In addition, we have the usual Poisson-equation for
$\Phi$, sourced by the total energy density (ie that of baryons and $\phi$ combined). 
Assuming spherical symmetry, the differential equation for $\phi$ can be solved analytically and the differential equation for $ \Phi $ can be readily numerically integrated. From the results one can then calculate
the total force acting on the baryons, which is a combination of the gravitational
pull (stemming from $\Phi$) and the phonon force (from $\phi$). Photons, on
the other hand, merely see the gravitational potential $\Phi$. 

There is however, an additional complication in this model, which is that at finite temperature 
the fluid will consist of two components,
one that is superfluid and one that behaves like a normal fluid.
To deal with this difficulty, we will follow here the procedure
proposed in \cite{Berezhiani2015, Berezhiani2017} and use the following modification of the zero-temperature Lagrangian (see also \cite{Nicolis:2011cs})
\beqn
{\cal L}_\phi = 2 \Lambda \frac{(2m)^{3/2}}{3} \chi \sqrt{|\chi - b \Upsilon|} ~, \label{Lphiups}
\eeqn
where
\beqn
\Upsilon := \mu - m \Phi + \dot \phi + \vec v \cdot \vec \nabla \phi  \,,
\eeqn
and $\vec{v} $ is the velocity vector of the normal fluid component.
As in Refs.~\cite{Berezhiani2015, Berezhiani2017} we will in the following work in the normal-fluid rest-frame, i.e. we take $ \vec{v} = 0 $.
Note that this choice of Lagrangian also cures a ghost in the zero-temperature model \cite{Berezhiani2015}.
We will use the Lagrangian (\ref{Lphiups}) up to the thermal radius, as laid out in \cite{Berezhiani2017},
and then match the superfluid's density profile to a standard {\sc NFW} profile \cite{Navarro:1996gj}.

We use the parameter values $b = 2$, $ \alpha = 5.7 $, $m = 1\,\textrm{eV} $, and $ \Lambda = 0.05\cdot10^{-3}\,\textrm{eV} $
as in \cite{Berezhiani2017}.
The chemical potential, $ \mu $, takes on different values for different galaxies as further discussed in Sec.~\ref{sec:calculation}.

With the so specified model, we can then calculate both the profile of the phonon field
 as well as the
gravitational potential from the initial conditions of the fields. This gives us
 the actual gravitational mass, consisting of both baryons and the energy-density of
$\phi$ -- which is what affects the trajectories of photons -- and we can also calculate the
total force acting on the baryons, composed of the gravitational force and the phonon force. 
Our procedure will then be to see whether we can find any initial conditions that fit the
data of the strong gravitational lenses.

\section{Data}

In Ref.~\cite{Berezhiani2017}, the effect of {\sc SFDM} on baryonic matter was tested by fitting rotation curves of two galaxies
(the low surface-brightness galaxy IC 2574 and the high surface-brightness galaxy UGC 2953).
Here, we are interested in the effect of {\sc SFDM} on both photons and baryons.
To this end, we will consider both the velocity dispersion and the Einstein radius of  
a set of 65 lenses from the Sloan Lens ACS ({\sc SLACS}) Survey. 
These lenses are classified as ellipticals and have velocity dispersions from {\sc SDSS} measurements as well as complete photometric data from Hubble Space Telescope (HST) measurements.
We have taken the redshifts, Einstein radii, effective radii, and velocity dispersions from Ref.~\cite{Auger2009} and the aperture radius and seeing from Ref.~\cite{Bolton2008}.

This is the same set of lenses that was previously studied in Ref.~\cite{Sanders2013} to evaluate the compatibility
of these lenses with {\sc MOND}. This previous study however only took into account the Einstein radius (and demonstrated
that fitting it can be achieved with {\sc MOND}). Our analysis improves on this previous work because we 
will investigate whether we can fit both the Einstein radius
{\sl and} the velocity dispersion at the same time to address the worry that this may not be possible if photons and
baryons feel different forces. 

The most interesting case would be to compare the gravitational mass $ M(r) $ inside some sphere with radius $ r $ to the mass inferred from kinematic measurements at the same radius.
However, with the measurements of the strong lensing systems from Ref.~\cite{Sanders2013} this is not possible for two reasons.
First, measurements of the Einstein radius $ R_\textrm{E} $ depend on the gravitational mass $ M_\textrm{E}(R_\textrm{E}) $ inside a cylinder along the line of sight rather than the mass inside a sphere with radius $ R_\textrm{E} $.
Second, for the galaxies considered here \cite{Bolton2008, Auger2009} only averaged velocity dispersions are available, hence we cannot
resolve a radial dependence.

Nevertheless, as a first check we should find out whether the Einstein radii and the velocity dispersions can be fit at the same time with {\sc SFDM}, and this will be our main task in the present work.

\section{Calculation}
\label{sec:calculation}

For our calculations, we follow Ref.~\cite{Sanders2013} in assuming spherical symmetry and a Jaffe mass distribution for the baryons \cite{Jaffe1983}.
Specifically, we take the baryonic energy density $ \rho_\textrm{b}(r) $ to be of the form
\begin{align}
 \label{eq:jaffe}
 \rho_\textrm{b}(r) = \frac{M_\textrm{b}}{4 \pi R_\textrm{J}} \frac{1}{r^2(1+r/R_\textrm{J})^2} \,.
\end{align}
Here, $ M_\textrm{b} $ is the total baryonic gravitational mass and $ R_\textrm{J} = 1.31\,R_\textrm{eff} $ is derived from the effective radius $ R_\textrm{eff} = \Theta_\textrm{eff} \cdot D_\textrm{l} $ with the angular effective radius $ \Theta_\textrm{eff} $ and the angular distance of the lens $ D_\textrm{l} $ \cite{Jaffe1983}.
For each galaxy, the effective radius of the Jaffe model is thereby determined from measurements.
This leaves the total baryonic mass, $ M_\textrm{b}, $ as the only free parameter of the baryonic mass distribution which is one of two free parameters we will use to fit the data.

As briefly laid out in section \ref{model}, the first step to calculate the Einstein radius and the velocity dispersion is solving the equations of motion of the phonon field $ \phi $ and the gravitational potential $ \Phi $.
This is described in detail in Sec.~V of Ref.~\cite{Berezhiani2017}.
Here, we only clarify a few points regarding the required initial conditions for the gravitational field.
With the analytic solution for $ \phi $, the resulting Poisson equation for $ \Phi $ inside the superfluid phase reads
\begin{align}
 \label{eq:poisson}
 \frac{1}{r^2}\, \partial_r \left(r^2 \partial_r \Phi(r) \right) = 4 \pi\, G \, \left[ \rho_b(r) + \rho_\textrm{SF}(a_\textrm{b}(r), \hat{\mu}(r)) \right] \,.
\end{align}
Here, $ a_b(r) $ is given by $ G\, M_\textrm{b}(r) / r^2 $ with the baryonic mass $ M_\textrm{b}(r) $ inside the sphere with radius $ r $, $ \rho_\textrm{SF} $ is the energy density of the superfluid, and
\begin{align}
  \label{eq:muhat}
 \hat{\mu}(r) \equiv \mu - m \, \Phi(r)\,.
\end{align}
Eq.~\ref{eq:poisson} can be rewritten so that it depends on $ \Phi $ only indirectly through $ \hat{\mu} $.
This is because we can use $ \hat{\mu}'(r) = -m \Phi '(r) $ on the left-hand side of Eq.~\ref{eq:poisson} (where a prime denotes a derivative with
respect to the radial coordinate $r$).
This means that given initial conditions $ \hat{\mu}(r_0) $ and $ \hat{\mu}'(r_0) $ for some $ r_0 $, the resulting 
equation can be solved numerically for $ \hat{\mu}(r) $.
With this procedure we can obtain $ \Phi'(r) $ (but not $ \Phi(r) $) without specifying $ \mu $ and $ \Phi(r_0) $ separately, 
only the combination $ \hat{\mu}(r_0) = \mu - m \, \Phi(r_0) $ is required.

In our calculations, we only need $ \Phi'(r) $ but not $ \Phi(r) $.
Therefore, we follow the procedure described in the previous paragraph and take the quantity $ \hat{\mu}(r_0) $ as the second free parameter in our calculation.
Since we will integrate the equations numerically, we take $ r_0 = 0.01\,\textrm{kpc} > 0 $ in order to avoid difficulties with
solving the Poisson equation at $ r = 0 $.
The other initial condition for the Poisson equation in spherical coordinates is usually $ \hat{\mu}'(r=0) = 0 $.
But since we solve the Poisson equation only for $ r \geq r_0 > 0 $, we instead take as our second initial condition $ \hat{\mu}'(r_0) = -m \cdot 4 \pi G r_0 (\rho_\textrm{b}(r_0) + \rho_\textrm{SF}(r_0)) $ which is obtained by expanding $ \hat{\mu}'(r) $ for small $ r $.

The next step consists in determining the radius at which the superfluid phase ends and the particle dark matter phase begins.
In Ref.~\cite{Berezhiani2017}, two algorithms were offered to determine this radius.
For simplicity, we will here take the thermalization radius $ R_{\rm T} $ as an estimate for the radius at which the superfluid phase ends.
According to Sec.~III of Ref.~\cite{Berezhiani2017} this radius is determined from
\begin{align}
 \label{eq:RT}
 \Gamma = t_\textrm{dyn}^{-1}\,, 
\end{align}
where $ \Gamma $ is the local self-interaction rate and $ t_\textrm{dyn} $ is the dynamical time.

We can estimate the thermal radius by using
 $ \Gamma = (\sigma / m) \, \mathcal{N} \, v \, \rho $, where $\sigma$ is the self-interaction rate, $ \mathcal{N} = (\rho/m) (2\pi/mv)^3 $
is the Bose-degeneracy factor, and $v$ is the average velocity of the particles. As in Ref.~\cite{Berezhiani2017}, we take 
$ \sigma/m = 0.01\,\textrm{cm}^2/\textrm{g} $.
Similarly, we can estimate $ t_\textrm{dyn} \approx r/v $ and $ v^2 \approx r \cdot \Phi'(r) $. Inserting these expressions
into (\ref{eq:RT}) allows one to obtain the thermalization radius $R_{\rm T}$. 

This
 procedure is arguably somewhat ad-hoc and good only approximately. Ideally one would want to be able to
 derive the composition of the two fluid-components as a function of radius (or pressure, respecively)
directly from the Lagrangian. We have checked, however, that our results do not depend much
 on the exact numerical factors in the above estimate and our general conclusion is not affected.
That the exact numerical factors do not affect our conclusion is supported by the analysis in 
Sec.~\ref{sec:nfwmatching}, where we have explicitly tried an alternative procedure for matching the superfluid core to an {\sc NFW} halo.

Further, following Refs.~\cite{Berezhiani2015, Berezhiani2017}, we assume that outside the superfluid phase, the
energy-density of the new field follows an {\sc NFW} profile.
For simplicity, we approximate this {\sc NFW} profile $ \rho_\textrm{NFW} $ outside the superfluid phase as $ \rho_\textrm{NFW} \propto 1/r^3 $
(though we checked that our results do not strongly depend on the exact form of the potential).

For the calculation of the angular Einstein radius $ \Theta_\textrm{E} $, we follow Ref.~\cite{Schwab2009} and fix the post-Newtonian parameter $ \gamma $ at its GR value $ \gamma \equiv 1 $ so that $ \Theta_\textrm{E} $ is determined by
\begin{align}
\label{eq:RE}
 \Theta_\textrm{E}^2 &= \frac{D_\textrm{ls}}{D_\textrm{l} D_\textrm{s}} \, 4 \, G \, M_\textrm{E}(R_\textrm{E}) \,, \\
 R_\textrm{E} &= D_\textrm{l} \, \Theta_\textrm{E} \,,
\end{align}
where $ G $ is Newton's gravitational constant, $ R_\textrm{E} $ is the Einstein radius, and $ D_\textrm{l} $, $ D_\textrm{s} $, and $ D_\textrm{ls} $ are the angular distances of the lens, the source, and the angular distance between the source and the lens, respectively.
Further, $ M_\textrm{E}(R_\textrm{E}) $ denotes the gravitational mass inside the cylinder with radius $ R_\textrm{E} $ along the line of sight.

For the calculation of the radial velocity dispersion $ \sigma_r $, we need to modify the formalism of Ref.~\cite{Schwab2009} due to the additional phonon force which acts on the baryons.
To this end, consider the formula for the velocity dispersion from Ref.~\cite{Schwab2009}
\begin{align}
 \label{eq:sigmar}
 \sigma_r^2(r) = \frac{G \int_r^\infty dr' \rho_\textrm{b}(r') M_\sigma(r') (r')^{2\beta-2}  }{ r^{2\beta} \rho_\textrm{b}(r) } \,,
\end{align}
with the anisotropy parameter $ \beta $, the baryonic energy density $ \rho_\textrm{b}(r) $ and the mass $ M_\sigma(r) $.
Here, $ M_\sigma(r) $ is calculated from the total acceleration of the baryons $ a_{\rm tot}(r) $ according to
\begin{align}
 \frac{G M_\sigma(r)}{r^2} = a_\textrm{tot}(r) \,.
\end{align}
We stress that $ a_\textrm{tot}(r) $ is the total acceleration which must include accelerations other than the gravitational acceleration if present.
This can be seen by rederiving this formula for $ \sigma_r $ following Ref.~\cite{Binney1980} in the case with forces other than the gravitational force acting on the baryons.

As already mentioned above, {\sc SFDM} predicts that $ a_\textrm{tot}(r) $ contains a contribution $ a_\phi(r) $ from the phonon force in addition to the usual gravitational acceleration $ a_\textrm{grav}(r) $:
 \begin{align}
 \label{eq:Msigma}
 \frac{G M_\sigma(r)}{r^2} = a_{\textrm{grav}}(r) + a_{\phi}(r) \, ,
 \end{align}
where $ a_\phi(r) $ is given by
\begin{align}
 a_\phi(r) = \alpha \frac{\Lambda}{m_\textrm{Pl}} \phi'(r) \,.
\end{align}

The measured velocity dispersion $ \sigma_* $ can then be obtained as in Ref.~\cite{Schwab2009} by using Eq.~\ref{eq:sigmar} for $ \sigma_r $ with the modified mass $ M_\sigma(r) $ from Eq.~\ref{eq:Msigma}:
\begin{align}
 \label{eq:sigma*}
 \sigma_*^2 &= \frac{\int_0^\infty dR\,R\,w(R) \int_{-\infty}^\infty dz \, \rho_\textrm{b}(r) \left(1 - \beta \frac{R^2}{r^2} \right) \sigma_r^2(r) }{\int_0^\infty dR\,R\,w(R) \int_{-\infty}^\infty dz \, \rho_\textrm{b}(r) } \,, \\
 w(R) &= e^{-R^2/2 \tilde{R}_\textrm{atm}^2}, \\
 \tilde{R}_\textrm{atm} &= \tilde{\sigma}_\textrm{atm} \, D_\textrm{l} \,, \\
 \tilde{\sigma}_\textrm{atm} &= \sigma_\textrm{atm} \sqrt{ 1+\theta_\textrm{ap}^2/4 + \theta_\textrm{ap}^4/40 } \,.
\end{align}
Here, $ r = \sqrt{R^2+z^2} $, $ \sigma_\textrm{atm} $ is the seeing, and $ \theta_\textrm{ap} $ is the spectrometric aperture.
In particular, we have  $ \sigma_\textrm{atm} = 1.4^{\prime\prime} $ and $ \Theta_\textrm{ap} = 1.5^{\prime\prime} $ \cite{Bolton2008}.
For simplicity, we take $ \beta = 0 $.

Angular distances $ D $ in the calculations for both the Einstein radius and the velocity dispersion are determined from the measured redshifts $ z_\textrm{l} $ and $ z_\textrm{s} $ of the lens and the source, respectively.
More conretely, we employ the relation
\begin{align}
 D(z_1, z_2) = \frac{1}{H(1+z_2)} \int_{z_1}^{z_2} \frac{dz'}{\sqrt{\Omega_\textrm{m}(1+z')^3+(1-\Omega_\textrm{m})}} \,,
\end{align}
where $ z_1 $ and $ z_2 $ are the redshifts of the objects whose angular distance is to be calculated.
Further, we take $ H = 70\,\textrm{km}\,\textrm{s}^{-1}\,\textrm{Mpc}^{-1} $ and $ \Omega_\textrm{m} = 0.3 $ following Ref.~\cite{Auger2009}.

\section{Fitting procedure}
\label{sec:fit}

We now search for values of the two parameters -- the baryonic mass $ M_{\rm b} $ and $ \hat{\mu}(r_0) $ -- that can reproduce both the measured Einstein radius and the measured velocity dispersion of each galaxy. This search proceeds as follows.

First, we fix $ M_{\rm b} $ at a starting value, for which we use the mass $ M_J $ listed in Ref.~\cite{Sanders2013}.
Next, we scan different values of $ \hat{\mu}(r_0) $, starting at $ 0.5 \cdot 10^{-5}\,\textrm{eV} $ until we find a value for which Eq.~\ref{eq:RT} has a solution and the calculated Einstein radius $ R_\textrm{E}^{\textrm{calc}} $ matches the measured Einstein radius $ R_\textrm{E}^{\textrm{meas}} $ to at least $ 0.01\,\textrm{kpc} $.
The step size of $ \hat{\mu}(r_0) $ starts at $ 0.3\cdot10^{-5}\,\textrm{eV} $ and is decreased each time $ R_\textrm{E}^\textrm{calc} - R_\textrm{E}^\textrm{meas} $ switches sign with a minimum step size of $ 0.005\cdot10^{-5}\,\textrm{eV} $.

If we do not find a value of $ \hat{\mu}(r_0) $ with $ |R_\textrm{E}^\textrm{calc} - R_\textrm{E}^\textrm{meas}| < 0.01\,\textrm{kpc} $, we decrease $ M_{\rm b} $ by $ 0.25\cdot10^{11}\,M_\odot $ and repeat the previous step.
We iterate this procedure until the measured Einstein radius is matched.

With the values of $ M_{\rm b} $ and $ \hat{\mu}(r_0) $ obtained in this way, we then calculate the velocity dispersion $ \sigma_*^\textrm{calc} $ and compare it to the measured value $ \sigma_*^\textrm{meas} $.
If $ \sigma_*^\textrm{calc} $ is closer to $ \sigma_*^\textrm{meas} $ than the error $ \sigma_*^\textrm{error} $ cited in Ref.~\cite{Auger2009}, we take this galaxy to be successfully fitted.
If not, we go back to the previous step, but instead of setting $ M_{\rm b} = M_J $ we increase or decrease $ M_{\rm b} $ by $ 0.25\cdot10^{11}\,M_\odot $ depending on whether $ \sigma_*^\textrm{meas} $ is larger or smaller than $ \sigma_*^\textrm{calc} $.

In this way, we obtain one value of $ M_\textrm{b} $ and $ \hat{\mu}(r_0) $ for each galaxy.
For some galaxies, these values correspond to a successful fit, for others they are the closest we could match the measured velocity dispersion so far.

We then proceed to scan values of $ M_\textrm{b} $ with a finer resolution.
In particular, we take the value for $ M_{\rm b} $ obtained in the previous steps and scan both larger and smaller values of $ M_\textrm{b} $ in steps of $ 0.05\cdot10^{11}\,M_\odot $.
For each value of $ M_{\rm b} $ we re-adjust $ \hat{\mu}(r_0) $ to fit the Einstein radius and then check whether or not the calculated velocity dispersion matches with the measured one.

For some galaxies, this procedure produces a list of values of $ M_\textrm{b} $ and $ \hat{\mu}(r_0) $ which fit both the Einstein radius as well as the velocity dispersion.
For galaxies for which we do not obtain a successful fit, we nevertheless record the values of $ M_\textrm{b} $ and $ \hat{\mu}(r_0) $ which give the closest match of the measured velocity dispersion.

Note that the above procedure only fully exploits the measurement errors in the velocity dispersion.
The Einstein radius is always matched to $ 0.01\,\textrm{kpc} $ which is usually less than the percent-level measurement error \cite{Bolton2008}.
Consequently, there may be additional values of $ M_{\rm b} $ and $ \hat{\mu}(r_0) $ which also fit some galaxies but which we do not find since we do not exploit the measurement errors on the Einstein radius.
However, we noticed that varying the Einstein radius by a few percent does not lead to qualitatively new results.
In particular, most parameters only vary by a few percent as a result of changing the Einstein radius by a few percent.
The only exception is the non-baryonic gravitational mass $ M_\textrm{DM} $ inside the virial radius $ r_{200} $.
As discussed below, this parameter is very sensitive to the other parameters so it is expected that it varies significantly when varying the Einstein radius.

It is important to note that for this procedure we require that a solution for Eq.~\ref{eq:RT} exists.
This means we enforce the existence of a superfluid phase in equilibrium.
This implies that if the parameters of a galaxy are such that they do not admit a 
superfluid phase in equilibrium, we will not be able to successfully fit this galaxy. Or, to put it
differently, if we cannot fit a galaxy it might mean this galaxy does not contain a superfluid phase.

We performed the calculation and the above described fitting procedure in Mathematica \cite{Mathematica}.

\section{Results}

\begin{figure}[t]
 \centering
 \includegraphics[width=.49\textwidth]{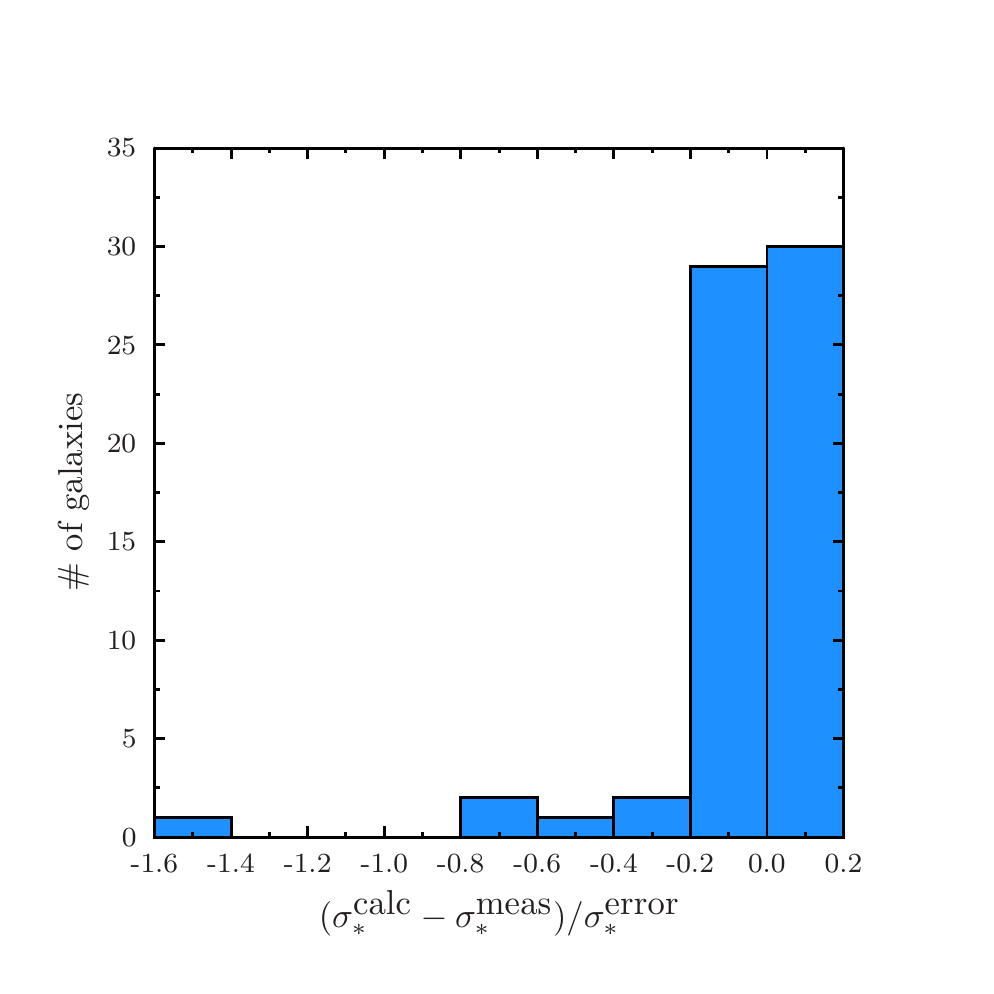}
 \includegraphics[width=.49\textwidth]{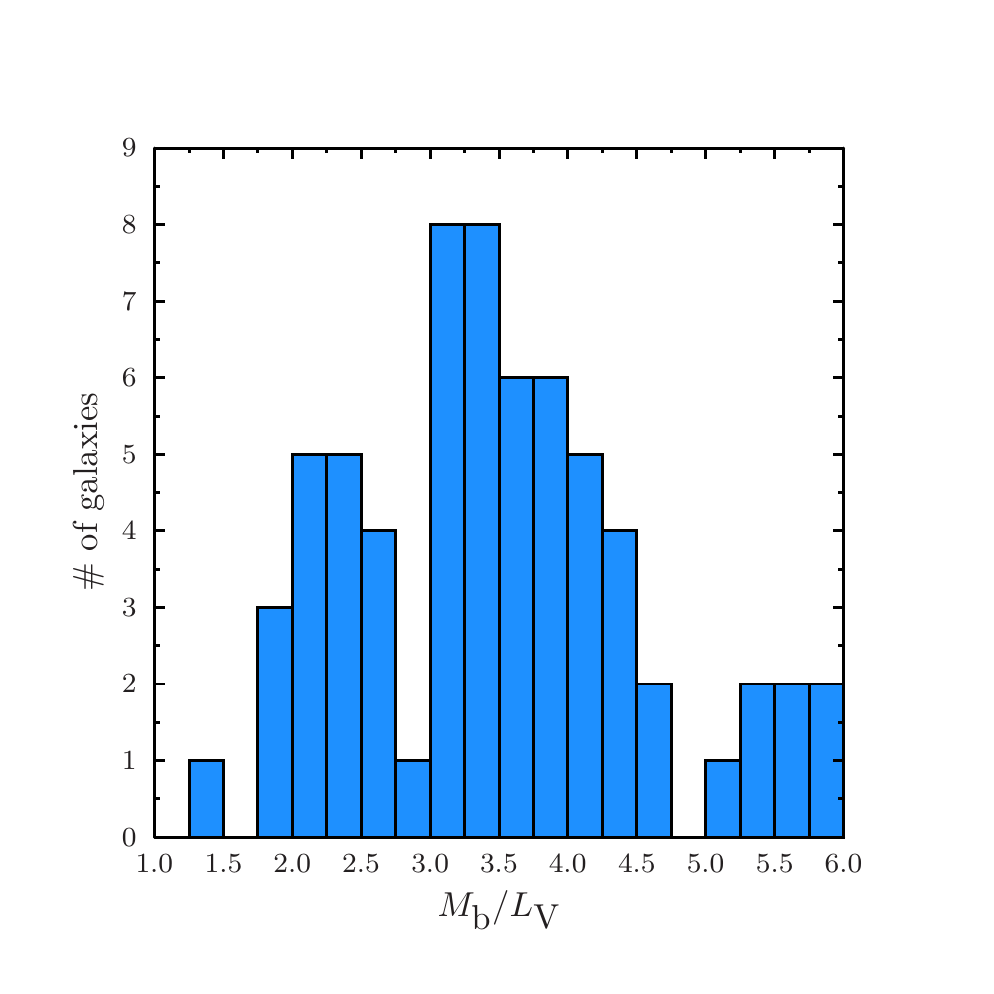}

 \caption{Left: Histogram of the $ (\sigma_*^\textrm{calc} - \sigma_*^\textrm{meas})/\sigma_*^\textrm{error} $ with minimum absolute value for each galaxy. Right: Histogram of the calculated stellar mass-to-light-ratios for $ M_\textrm{b} $ and $ \hat{\mu}(r_0) $ such that $ |\sigma_*^\textrm{calc} - \sigma_*^\textrm{meas}| $ is minimal.}

 \label{fig:histogramsa}
\end{figure}

\begin{figure}[t]
 \centering
 \includegraphics[width=.49\textwidth]{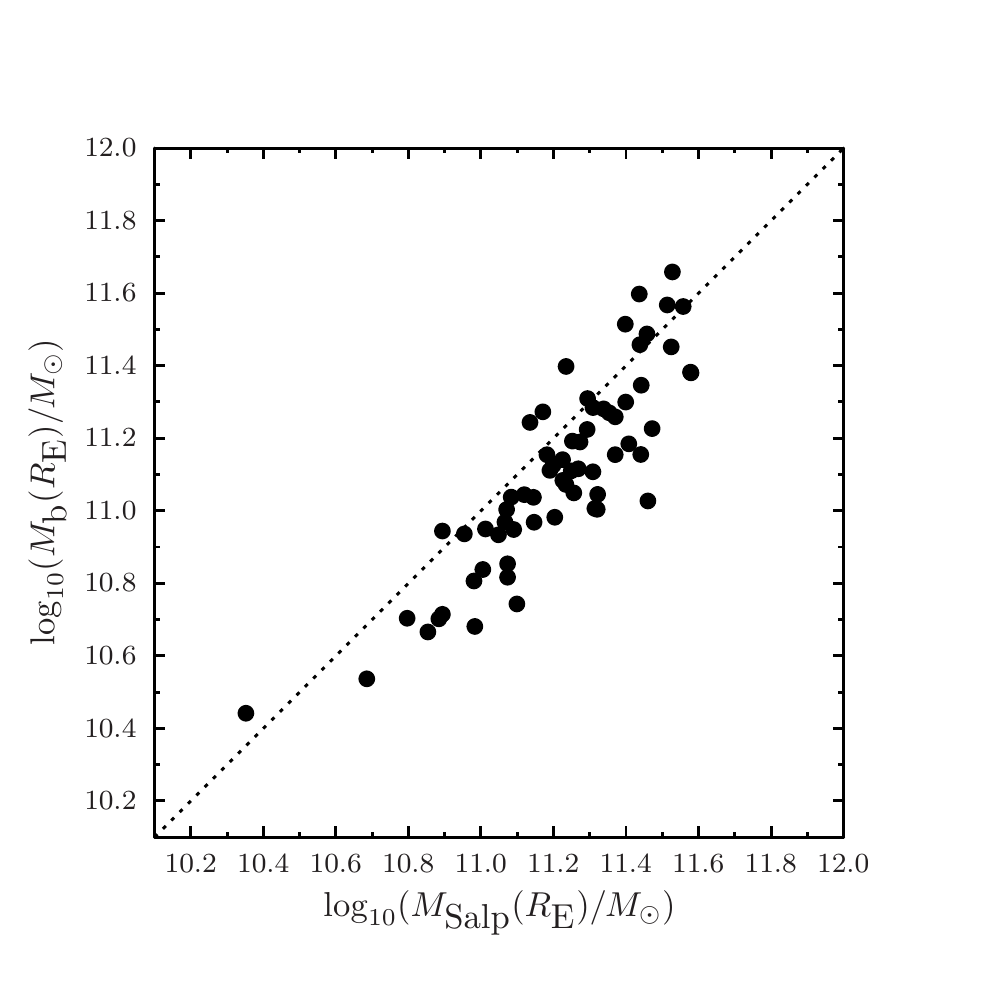}
 \includegraphics[width=.49\textwidth]{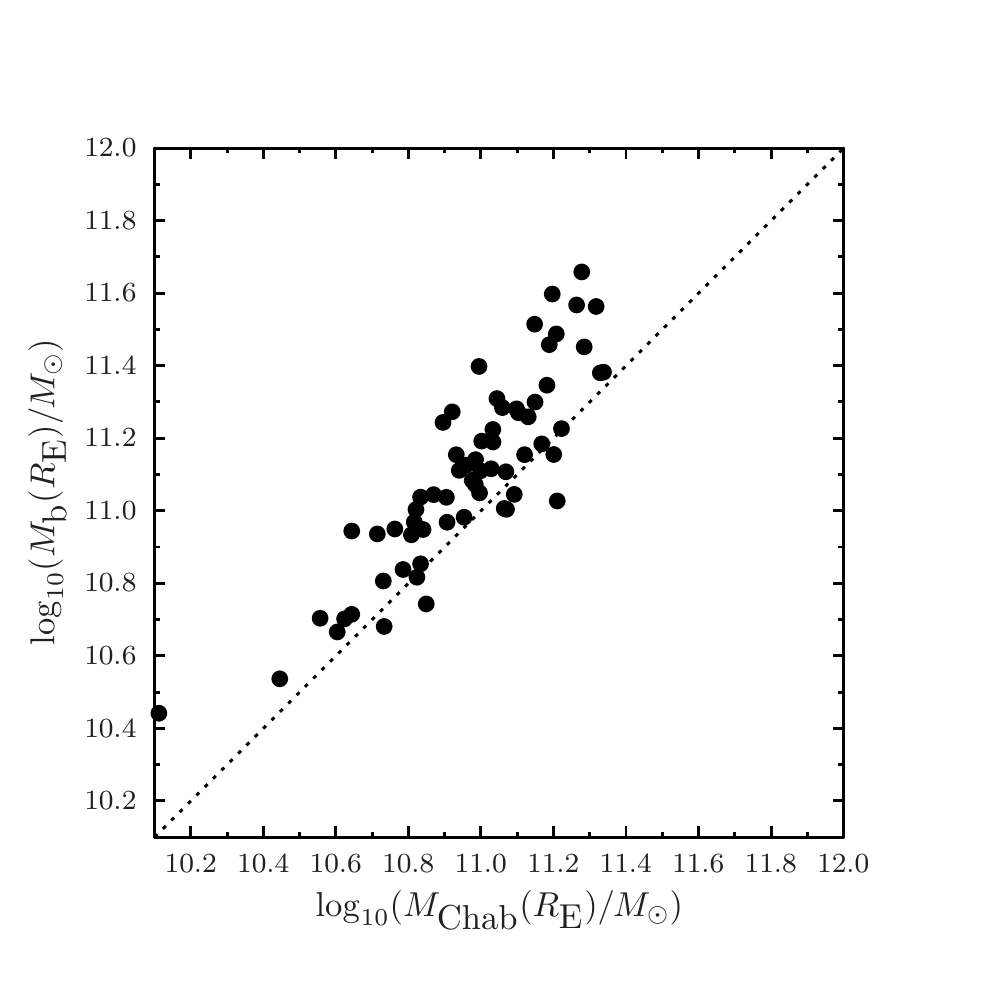}
 \caption{Left: Stellar mass within the cylinder with radius $ R_\textrm{E}^\textrm{meas} $ as calculated with the stellar mass estimated from the Salpeter IMF compared to that calculated with our $ M_\textrm{b} $. Right: The same as left, but with the Chabrier IMF instead of the Salpeter IMF.}
 \label{fig:MIMFvsMb}
\end{figure}

\begin{figure}[t]
 \centering
 \includegraphics[width=.49\textwidth]{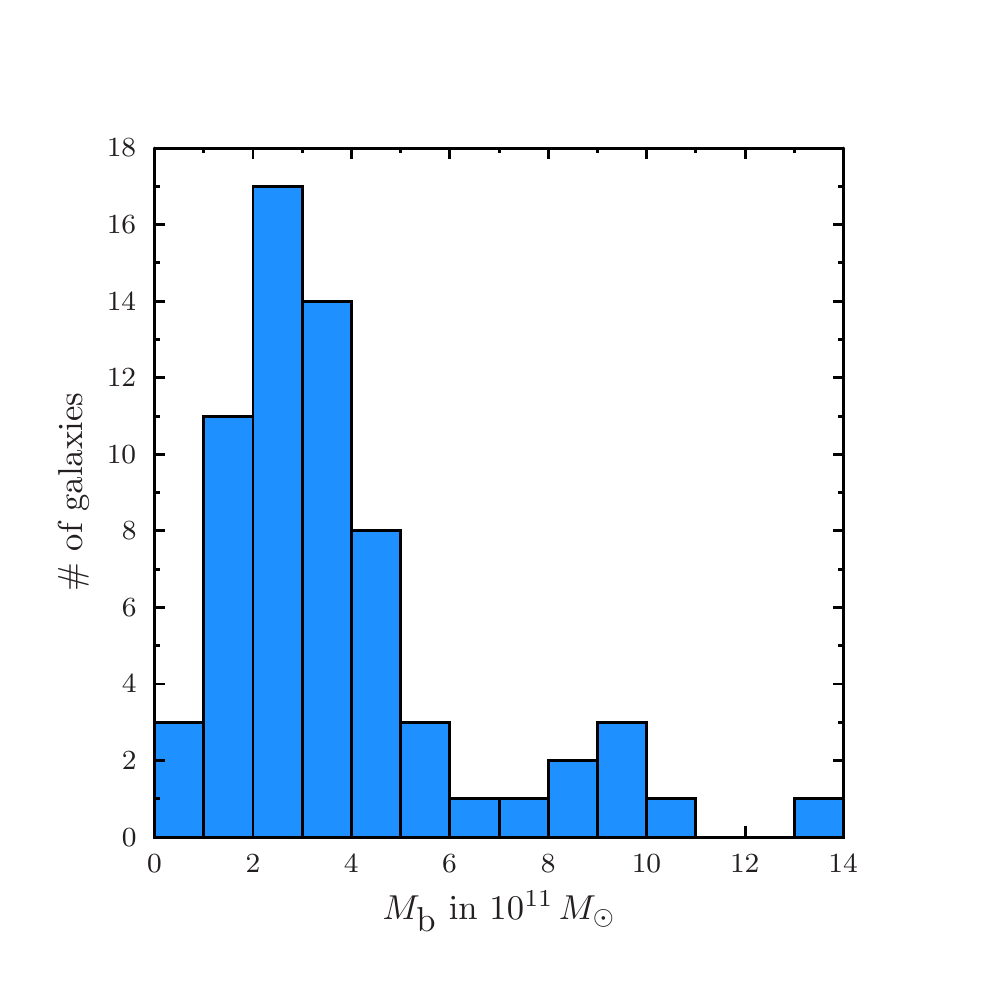}
 \includegraphics[width=.49\textwidth]{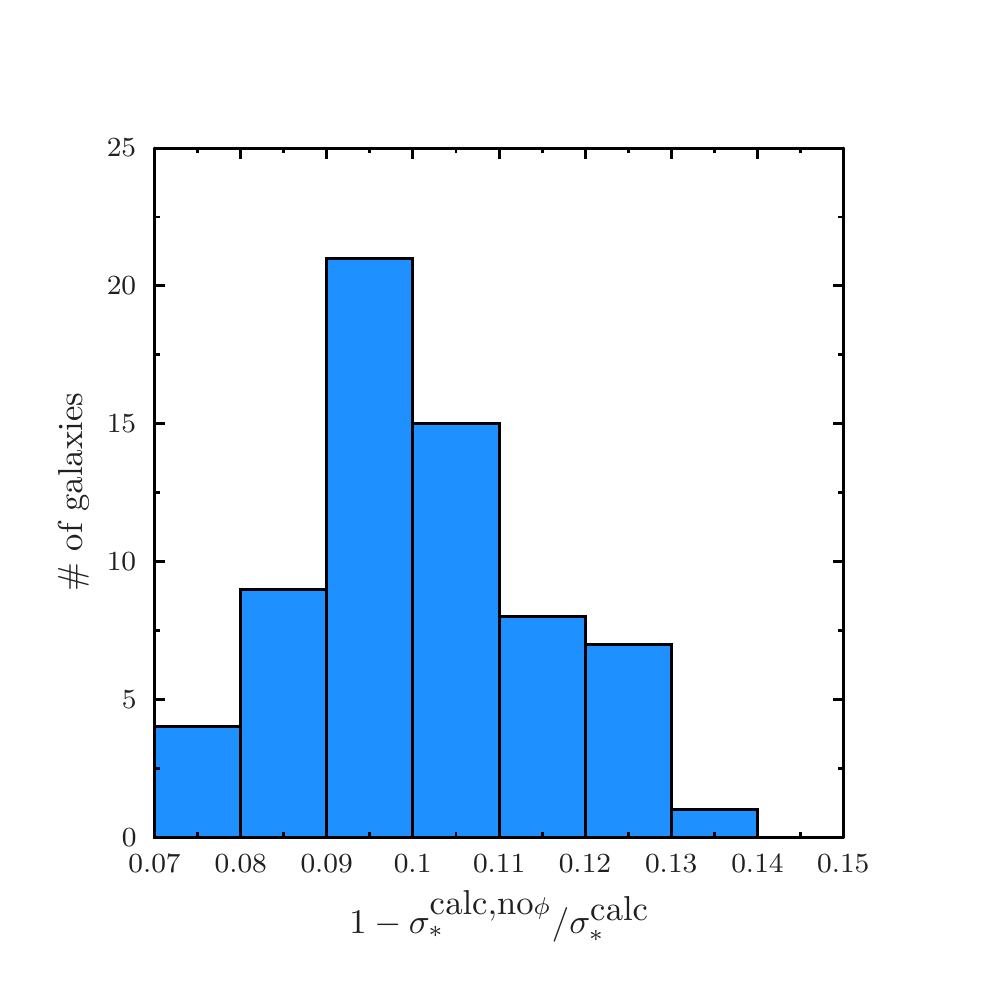}

 \caption{Left: Histogram of the total baryonic masses which lead to minimal $ |\sigma_*^\textrm{calc} - \sigma_*^\textrm{meas}| $ for each galaxy. Right: Histogram of the fractions of $ \sigma_*^\textrm{calc} $ due to the phonon force for $ M_\textrm{b} $ and $ \hat{\mu}(r_0) $ such that $ |\sigma_*^\textrm{calc} - \sigma_*^\textrm{meas}| $ is minimal.}
 \label{fig:histogramsb}
\end{figure}

The results of our calculation are summarized in Table~\ref{tab:results} and Figs.~\ref{fig:histogramsa}, \ref{fig:MIMFvsMb},  and \ref{fig:histogramsb}.
It can be seen from Fig.~\ref{fig:histogramsa}, left, that we are able to successfully fit 64 out of 65 galaxies.
The stellar mass-to-light ratios, $ M/L_\textrm{V} $, of the fitted galaxies are reasonable, see Fig.~\ref{fig:histogramsa}, right.
They are generally somewhat lower than those found in Ref.~\cite{Sanders2013} for MOND:
We find an averaged $ M/L_\textrm{V} $ of $ 3.5\pm1.1 $ compared to the $ 4.2\pm1.0 $ obtained in Ref.~\cite{Sanders2013}.

Assuming a Salpeter and a Chabrier initial mass function ({\sc IMF}), Ref.~\cite{Auger2009} gives an estimate of the total stellar mass of each galaxy which we can compare to the $ M_\textrm{b} $ obtained from our fitting procedure.
To this end, we follow Ref.~\cite{Sanders2013} and use our Jaffe mass model with the measured effective radius to calculate the stellar mass inside the measured Einstein radius.
This can be done for our fitted $ M_\textrm{b} $ as well as for the stellar mass estimate from each IMF.
The results are shown in Fig.~\ref{fig:MIMFvsMb}.
It can be seen that the Salpeter {\sc IMF} tends to give stellar masses larger than our $ M_\textrm{b} $, while the Chabrier {\sc IMF} tends to give stellar masses smaller than our $ M_\textrm{b} $.
This again shows that our fitted stellar masses are reasonable.

\subsection{Contribution of phonon force }
\label{sec:phonon}

To find out how much the phonon force contributes to $ \sigma_*^\textrm{calc} $, we have also calculated a velocity dispersion $ \sigma_*^{\textrm{calc,no}\phi} $ with $ a_\phi \equiv 0 $.
As can be seen from Fig.~\ref{fig:histogramsb}, right, the phonon force contributes around 10\% to the calculated velocity dispersion $ \sigma_*^\textrm{calc} $.

At first sight it might seem surprising that the phonon force does not make a larger contribution to $ \sigma_*^\textrm{calc} $ since the phonon force is responsible for the {\sc MOND}-like behavior discussed in Refs.~\cite{Berezhiani2015, Berezhiani2017}. One therefore might expect that it
should dominate over the gravitational force at least in a good part of each galaxy.

In order to understand the small contribution of the phonon force to $ \sigma_*^\textrm{calc} $, we can
approximate the phonon field $ \phi'(r) $ by the expression it takes in the {\sc MOND}-regime \cite{Berezhiani2015}
\begin{align}
 \label{eq:phimond}
 \phi'(r) = \sqrt{\frac{\alpha^3 \Lambda^2}{m_\textrm{Pl}} \frac{G M_{\rm b}(r)}{r^2}} \,.
\end{align}
We can neglect the contributions of non-baryonic gravitational matter for reasons discussed in Sec.~\ref{sec:nonbaryonic}.
From this, we can then derive analytic expressions for the contribution $ \sigma_{r, \phi}^2(r) $ of the phonon force to $ \sigma_{r}^2(r) $  and for the contribution $ \sigma_{r, \textrm{b}}^2(r) $ of the gravitational force due to the baryons. The resulting
expressions however are not very illuminating, so we will instead look at an example to see what is going on.

\begin{figure}[th]
{\caption{Different contributions to $ \sigma^2_r(r) $ for the galaxy J0029-0055 inside the superfluid phase up to the galaxy's thermal radius $ R_\textrm{T} \approx 153\,\textrm{kpc} $.
 Shown is the phonon force approximated by its {\sc MOND}-regime form $ \sigma^2_{r,\phi}(r) $ and the contribution due to the gravitational pull of the baryons $ \sigma^2_{r,b}(r) $. } \label{fig:sigmar}}
 {
\includegraphics[width=.9\textwidth]{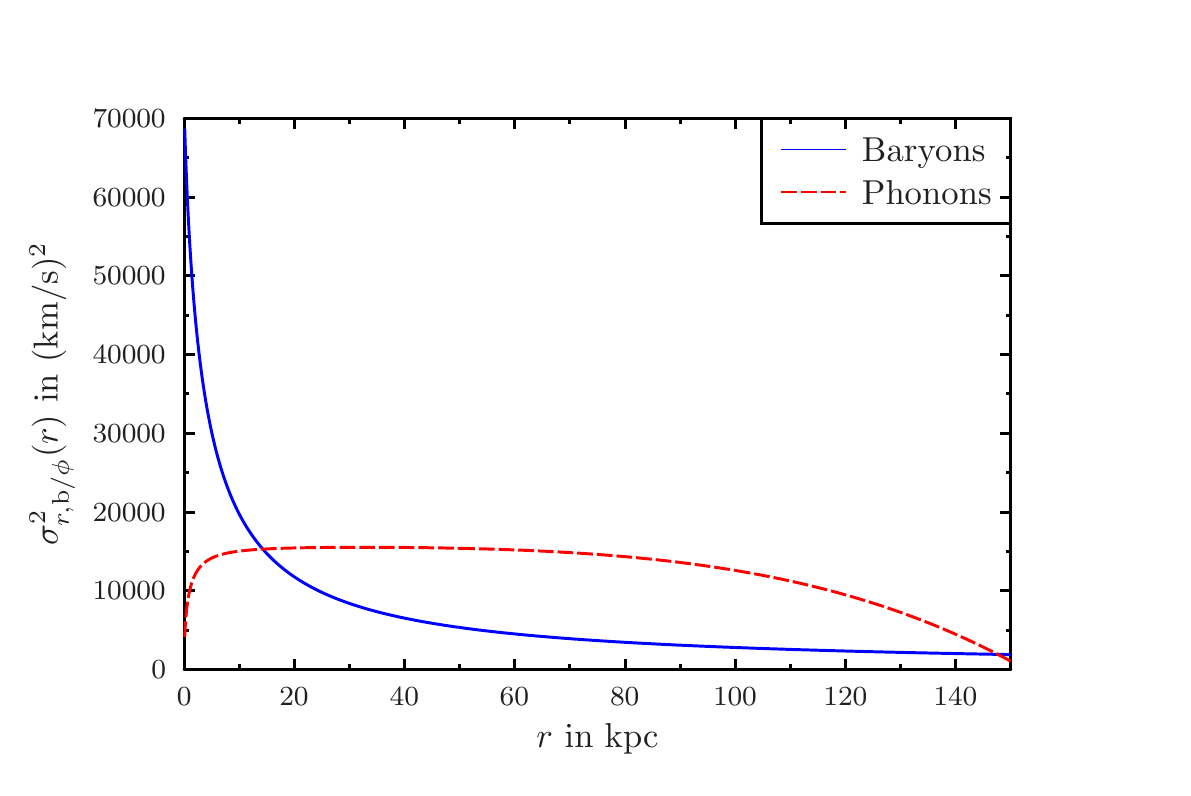}}
\end{figure}

Fig.~\ref{fig:sigmar} shows both $ \sigma_{r,\rm{b}}^2(r) $ and $ \sigma_{r, \phi}^2(r) $ for the galaxy J0029-0055.
It can be seen that for most of the superfluid phase the phonon force is the dominant contribution to $ \sigma_r^2(r) $.
However, this is not true for small radii.
Indeed, we find for $ r \ll R_\textrm{J} $:
\begin{align}
 \label{eq:sigmarapprox}
 \sigma_{r, \rm{b}}^2(r) &\approx \frac{G M_\textrm{b}}{2 R_\textrm{J}^2} \left( R_\textrm{J} - 4 r \right) \,, \\
 \sigma_{r,\phi}^2(r) &\approx \frac23 \alpha^{3/2} \Lambda \sqrt{\frac{G M_\textrm{b}}{m_\textrm{Pl} R_\textrm{J}} r} \,.
\end{align}
By plugging in typical numbers, one can confirm that in this regime the phonon force is usually smaller than the gravitational 
force excerted merely by the baryons.

In more detail, the reason for the relatively small contribution of the phonon force to $ \sigma_*^\textrm{calc} $ is that $ \sigma_*^\textrm{calc} $ is dominated
by the contributions to $ \sigma_r(r) $ that come from small $ r $.
According to Eq.~\ref{eq:sigma*}, $ (\sigma_*^\textrm{calc})^2 $ is proportional to the weighted cylindrical integral over $ \sigma_{r}^2(\sqrt{R^2+z^2}) $ with a weight factor proportional to $ R \cdot \exp(-R/R_\textrm{atm}) \cdot \rho_\textrm{b}(\sqrt{R^2+z^2}) $.
Both the exponential with $ R_\textrm{atm} \sim 6\,\textrm{kpc} $ and the fall-off of $ \rho_\textrm{b} $ lead to $ (\sigma_*^\textrm{calc})^2 $ being mostly calculated from $ \sigma_{r}^2(r) $ at radii where $ \sigma^2_{r,\phi}(r) $ is small compared to $ \sigma^2_{r,\textrm{b}}(r) $.
Thus, the phonon force does not dominate the final result for $ \sigma_*^\textrm{calc} $.

We want to emphasize that this finding agrees with the more general argument given in \cite{Sanders2013} that strong gravitational
lensing is not sensitive to the distinction between {\sc MOND} and cold dark matter because it mainly probes the
mass in the galactic center which is dominated by baryons either which way. At small radii, the baryonic acceleration 
is larger than the {\sc MOND}-acceleration scale $ a_0 $.
Therefore, the force acting on the baryons should mainly be the gravitational force due to the baryons themselves.

\subsection{Contribution of non-baryonic gravitational mass}
\label{sec:nonbaryonic}

Next, we would like to discuss how the non-baryonic gravitational mass contributes to the Einstein radius and the velocity dispersion.

First, consider the velocity dispersion.
From Eq.~\ref{eq:sigma*} and Eq.~\ref{eq:sigmar} and leaving out the phonon force which was already discussed in Sec.~\ref{sec:phonon}, we have
\begin{align}
 \label{eq:sigmacompare}
 \sigma_*^2 &\propto \int_0^\infty dR\,R\,w(R) \int_{-\infty}^\infty dz \, \int_{\sqrt{R^2+z^2}}^\infty dr \, \rho_\textrm{b}(r) \left( M_b(r) + M_\textrm{DM}(r) \right) \,, \\
 M_\textrm{b}(r) &\equiv \int_0^r dr' 4\pi \,r'^2\, \rho_\textrm{b}(r') \,, \\
 M_\textrm{DM}(r) &\equiv \int_0^r dr' 4\pi \,r'^2 \, \rho_\textrm{DM}(r') \,.
\end{align}
Here, $ \rho_\textrm{DM}(r) $ is the same as $ \rho_\textrm{SF}(r) $ for $ r < R_\textrm{T} $ and is proportional to $ 1/r^3 $ for $ r > R_\textrm{T} $.
Eq.~\ref{eq:sigmacompare} implies that the relative size of the baryonic and the non-baryonic contributions to $ \sigma_* $ is determined by the relative size of $ M_\textrm{b}(r) $ and $ M_\textrm{DM}(r) $, i.e. the baryonic and the non-baryonic mass inside a sphere with radius $ r $.

In contrast to this, the relative size of the baryonic and the non-baryonic contributions to the Einstein radius $ R_\textrm{E} $ is determined by the relative size of the baryonic and the non-baryonic mass inside a cylinder with radius $ R_\textrm{E} $.
More precisely, according to Eq.~\ref{eq:RE}:
\begin{align}
 R_\textrm{E}^2 &\propto M_\textrm{E,b}(R_\textrm{E}) + M_\textrm{E,DM}(R_\textrm{E}) \,, \\
 M_\textrm{E,b}(R) &\equiv \int_0^R dR' \int_{-\infty}^{\infty} dz\,  2\pi \, R' \, \rho_\textrm{b}(\sqrt{R'^2+z^2}) \,, \\
 M_\textrm{E,DM}(R) &\equiv \int_0^R dR' \int_{-\infty}^{\infty} dz\,  2\pi \, R' \, \rho_\textrm{DM}(\sqrt{R'^2+z^2}) \,.
\end{align}

We will now argue that the non-baryonic mass is negligible for the calculation of the velocity dispersion, but does affect the Einstein radius.

\begin{figure}[p]
 \vspace*{-2cm}
 \includegraphics[width=.9\textwidth]{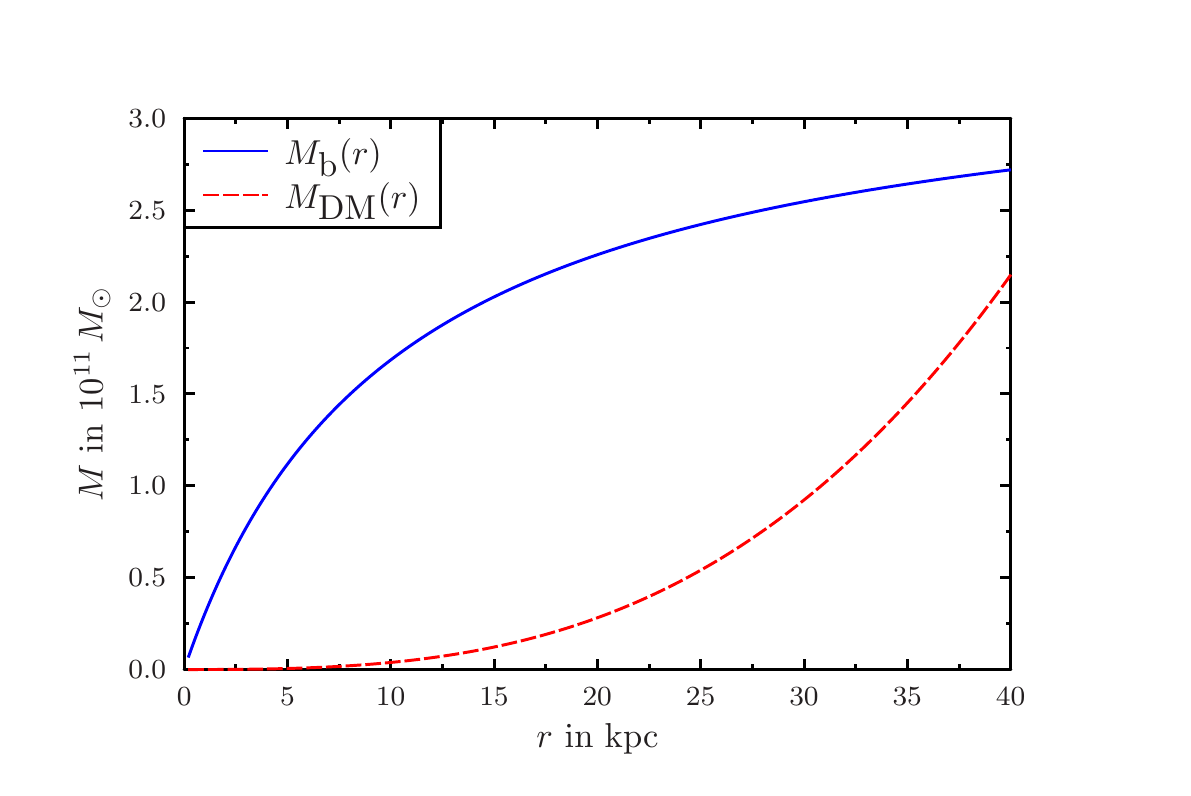}  
 \vspace*{-1cm}
 
 \includegraphics[width=.9\textwidth]{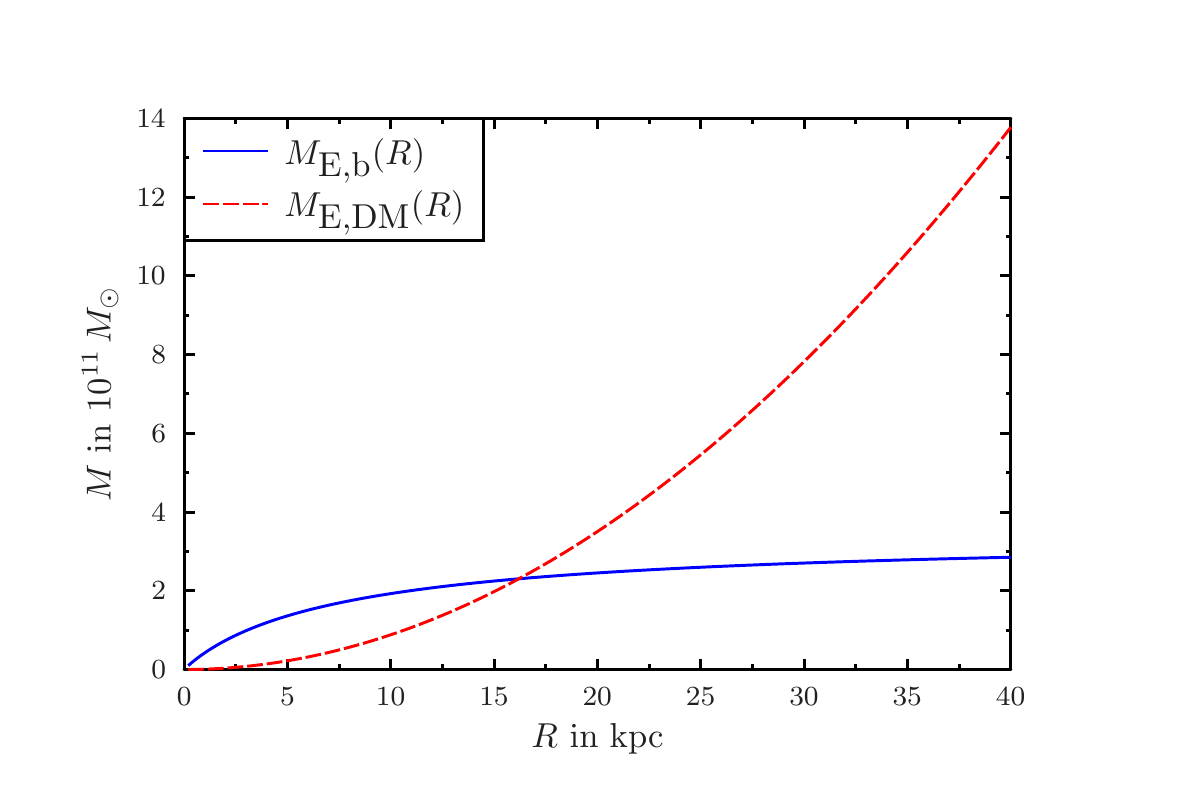}
 
 \caption{Top: The baryonic and non-baryonic gravitational masses $ M_\textrm{b}(r) $ and $ M_\textrm{DM}(r) $ inside a sphere with radius $ r $ for the galaxy J0029-0055.
Bottom: The non-baryonic gravitational masses $ M_\textrm{E,b}(R) $ and $ M_\textrm{E,DM}(R) $ inside a cylinder with radius $ R $ for the galaxy J0029-0055.}
 \label{fig:MbMDM}
\end{figure}

Regarding the velocity dispersion, note that the baryonic mass is negligible for large spherical radii since it is typically concentrated inside a few $ 10\,\textrm{kpc} $ for the galaxies considered here.
Similarly, for small spherical radii, the mass of the superfluid halo is negligible compared to the baryonic mass\footnote{This can be checked explicitly by using the MOND-regime form of $ \phi' $ from Eq.~\ref{eq:phimond} in $ \rho_\textrm{SF}(r) = 2/3 m^2 \Lambda \left( 6 m \hat{\mu}(r) + \phi'(r)^2 \right) / \sqrt{ 2 m \hat{\mu}(r) + \phi'(r)^2} $ with values of $ \hat{\mu} $ close to their initial values listed in Table~\ref{tab:results}.} as illustrated in Fig.~\ref{fig:MbMDM}, top, for the galaxy J0029-0055.
Since the velocity dispersion is mainly calculated from small spherical radii as discussed in Sec.~\ref{sec:phonon}, it follows that the non-baryonic mass can be neglected when calculating $ \sigma_* $.

For the calculation of the Einstein radius, both small and large spherical radii contribute since the Einstein radius is determined by the mass inside a cylinder with radius $ R_\textrm{E} $.
This is especially important for the non-baryonic matter $ M_\textrm{E,DM}(R_\textrm{E}) $ since there is much more non-baryonic matter than baryonic matter in total, but most of this non-baryonic matter is located at large spherical radii.
This is illustrated in Fig.~\ref{fig:MbMDM}, bottom.
As a result, the non-baryonic matter $ M_\textrm{E,DM}(R_\textrm{E}) $ is usually non-negligible for the calculation of the Einstein radius.
In particular, for the galaxy J0029-0055, we have $ M_\textrm{E,DM}(R_\textrm{E}) / M_\textrm{E,b}(R_\textrm{E}) \approx 10\,\% $.

\subsection{Sensitivity of $ M_\textrm{DM} $ to initial condition}
\label{sec:mdmsensitivity}

In this subsection, we will comment on one peculiarity in our results.
Namely, the total non-baryonic gravitational mass $ M_\textrm{DM} $ is quite sensitive to the parameters $ \hat{\mu}(r_0) $ and $ M_\textrm{b} $.
This regards the non-baryonic gravitational mass in the superfluid phase $ M_\textrm{DM}^\textrm{SF} $ as well as that outside the superfluid phase $ M_\textrm{DM}^\textrm{NFW} $.
For example, for the galaxy J0029-0055 the ratio of the maximum value $ M_\textrm{DM}|_\textrm{max} $ of $ M_\textrm{DM} $ to the respective minimum value $ M_\textrm{DM} |_\textrm{min} $ is about $ 200 $.
This is despite the fact that the values of $ \hat{\mu}(r_0) $ corresponding to the  maximum and minimum $ M_\textrm{DM} $ differ only by a factor of about $ 1.6 $ and the values of $ M_\textrm{b} $ corresponding to the maximum and minimum $ M_\textrm{DM} $ differ only by a factor of about $ 0.8 $.
We will now try to understand where this sensitivity comes from.

Numerically, for the galaxy J0029-0055, the differences in $ \hat{\mu}(r_0) $ and $ M_\textrm{b} $ imply a factor of about $ 2.9 $ in the thermal radius $ R_\textrm{T} $ and a factor of $ 3.3 $ in the average value of the non-baryonic energy density inside the superfluid phase $ \rho_\textrm{SF} $.
Thus, those quantities do not seem to be particularly sensitive to $ M_\textrm{b} $ and $ \hat{\mu}(r_0) $.

Let us now consider the non-baryonic gravitational mass in the superfluid phase $ M_\textrm{DM}^\textrm{SF} $.
For our purposes, it suffices to approximate $ \rho_\textrm{SF} $ as constant which gives
\begin{align}
 \label{eq:sfmass}
 M_\textrm{DM}^\textrm{SF} \sim \frac{4 \pi}{3} \, R_\textrm{T}^3 \, \rho_\textrm{SF} \,.
\end{align}
With the numerical values cited above, this implies a ratio between the maximum and minimum values of $ M_\textrm{DM}^\textrm{SF} $ of about $ 80 $.
The actual value from the full calculation is $ 120 $ which is of the same order of magnitude.

A similar estimate is possible for the non-baryonic gravitational mass outside the superfluid phase $ M_\textrm{DM}^\textrm{NFW} $ if we approximate the virial radius $ r_{200} $ as the radius at which the non-baryonic energy density drops below $ \rho_{200} \equiv 200 \cdot 3 \, H^2/(8 \pi G) $.
Here, $ H $ is the Hubble rate.
With this we obtain:
\begin{align}
 \label{eq:nfwmass}
 M_\textrm{DM}^\textrm{NFW} \sim \frac{4 \pi}{3} R_\textrm{T}^3 \, \rho_\textrm{SF}(R_\textrm{T}) \, \ln\left(\frac{\rho_\textrm{SF}(R_\textrm{T})}{\rho_{200}}\right) \,.
\end{align}
With $ R_\textrm{T} $ as cited above and using $ \rho_\textrm{SF}(R_\textrm{T})|_\textrm{min} = 1.2\cdot10^{-26}\,\textrm{g}/\textrm{cm}^3 $ as well as $ \rho_\textrm{SF}(R_\textrm{T})|_\textrm{max} = 5.9 \cdot \rho_\textrm{SF}(R_\textrm{T})|_\textrm{min} $, this implies a ratio of the maximum and minimum values of $ M_\textrm{DM}^\textrm{NFW} $ of about $ 280 $.
The actual value from the full calculation is $ 230 $ which is again of the same order of magnitude.

Thus, while $ R_\textrm{T} $ and $ \rho_\textrm{SF} $ are not particularly sensitive to the values of $ \hat{\mu}(r_0) $ and $ M_\textrm{b} $, Eq.~\ref{eq:nfwmass} and Eq.~\ref{eq:sfmass} show that the total non-baryonic gravitational mass is indeed quite sensitive to these values.

\begin{figure}[t]
 \centering
 \includegraphics[width=.9\textwidth]{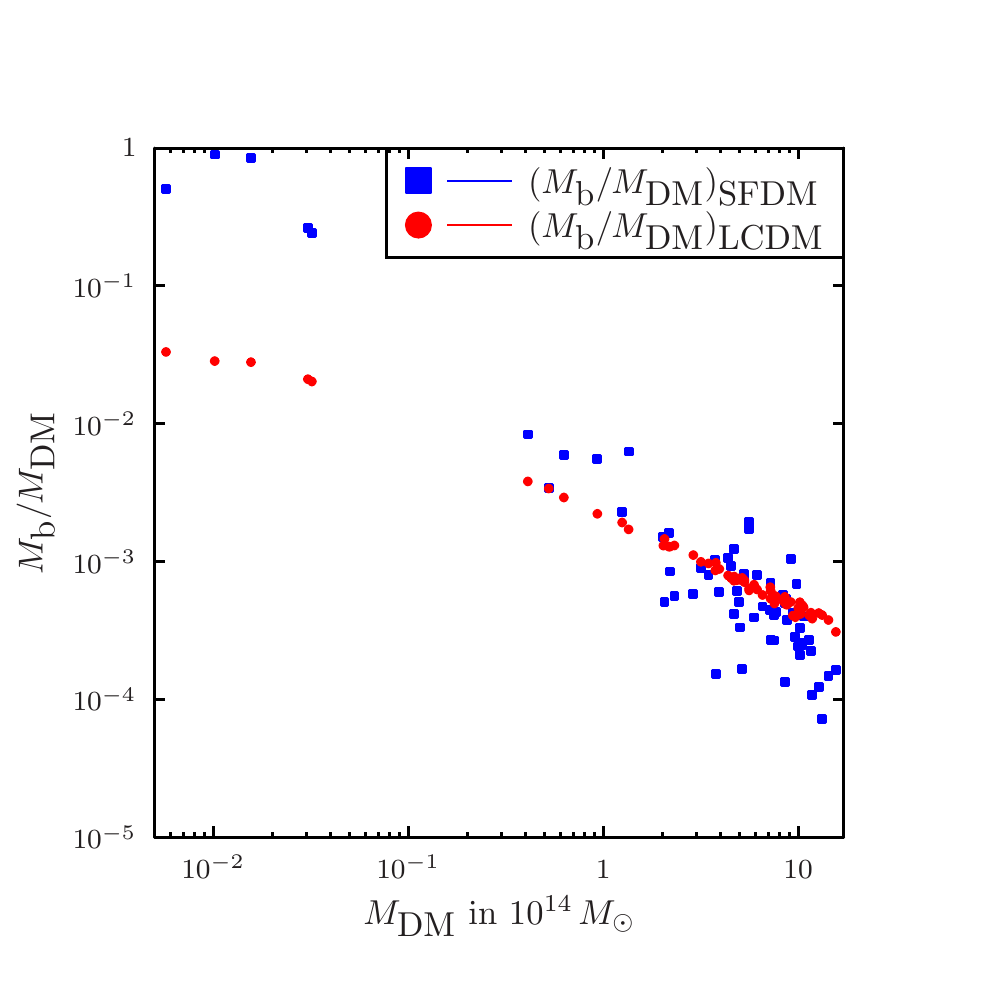}
 \caption{The ratio of stellar and halo mass over the halo mass, both for our calculations and for the $\Lambda$CDM mass-concentration relation given our $ M_\textrm{DM} $.}
 \label{fig:MboverMDM}
\end{figure}

So far, it is not well-understood how the cosmological case works with  {\sc SFDM}, so that it is not clear if our values for $ M_\textrm{DM} $ would fit the cosmological data.
Nevertheless, it is interesting to compare our results to $\Lambda$CDM abundance matching expectations.
In Fig.~\ref{fig:MboverMDM}, we show our results for $ M_\textrm{b} / M_\textrm{DM} $ together with the 
$ \Lambda$CDM expectations determined by the following equation from Ref.~\cite{Moster2013}:
\begin{align}
 \left( \frac{M_\textrm{b}}{M_\textrm{DM}} \right)_{\Lambda \textrm{CDM}} = 2\,N\,\left[ \left(\frac{M_\textrm{DM}}{M_1} \right)^{-\beta} + \left( \frac{M_\textrm{DM}}{M_1} \right)^\gamma \right]^{-1} \,.
\end{align}
Here, $ N $, $ M_1 $, $ \beta $, and $ \gamma $ are redshift-dependent numbers which we take from Eqs. (11)-(14) and Table 1 of Ref.~\cite{Moster2013}.
We see that most of our galaxies have $ M_\textrm{DM} \gtrsim 10^{14}\,M_\odot $ and the corresponding values of $ M_\textrm{b}/M_\textrm{DM} $ are very roughly compatible with the $\Lambda$CDM values.
In contrast, for galaxies with $ M_\textrm{DM} \lesssim 10^{13} M_\odot $ we obtain significantly higher $ M_\textrm{b}/M_\textrm{DM} $ than $ \Lambda$CDM. While this is interesting, we cannot draw any further conclusion from this without a
cosmological model for {\sc SCDM}.

\begin{figure}[t]
 \centering
 \includegraphics[width=.49\textwidth]{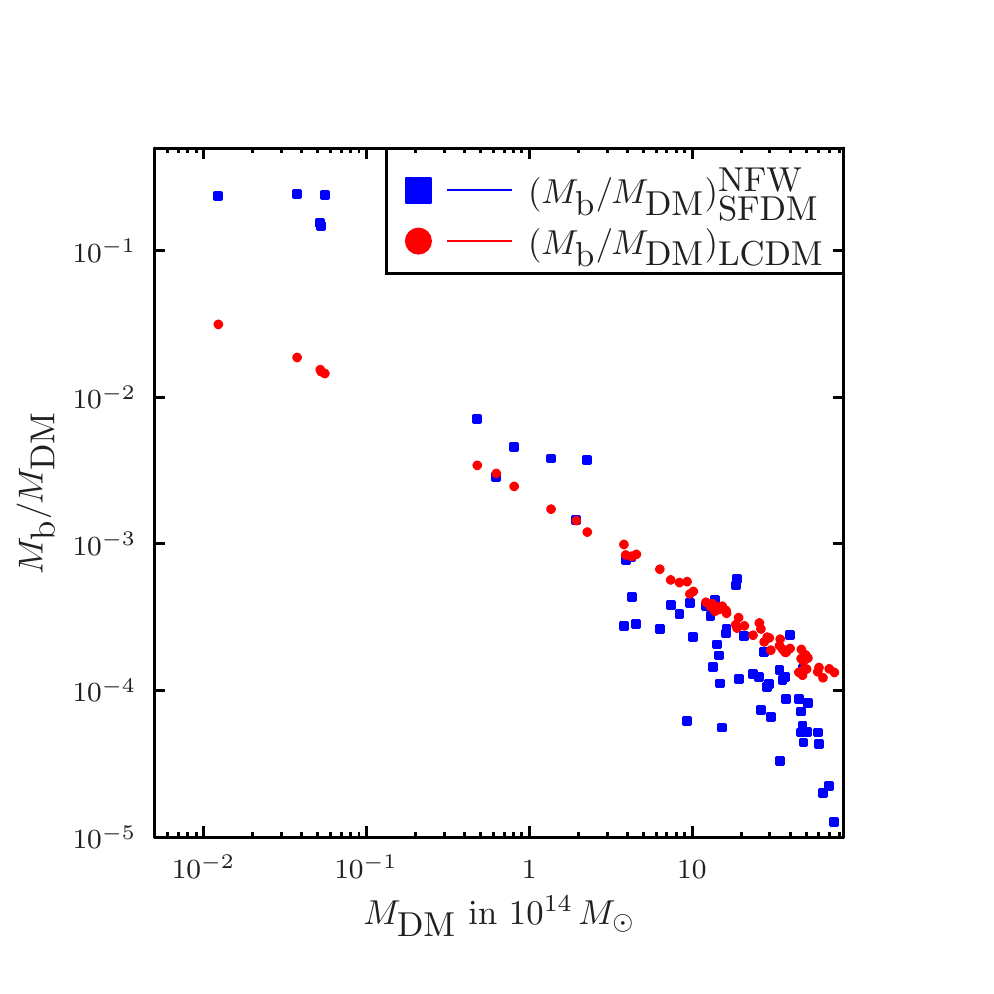}
 \includegraphics[width=.49\textwidth]{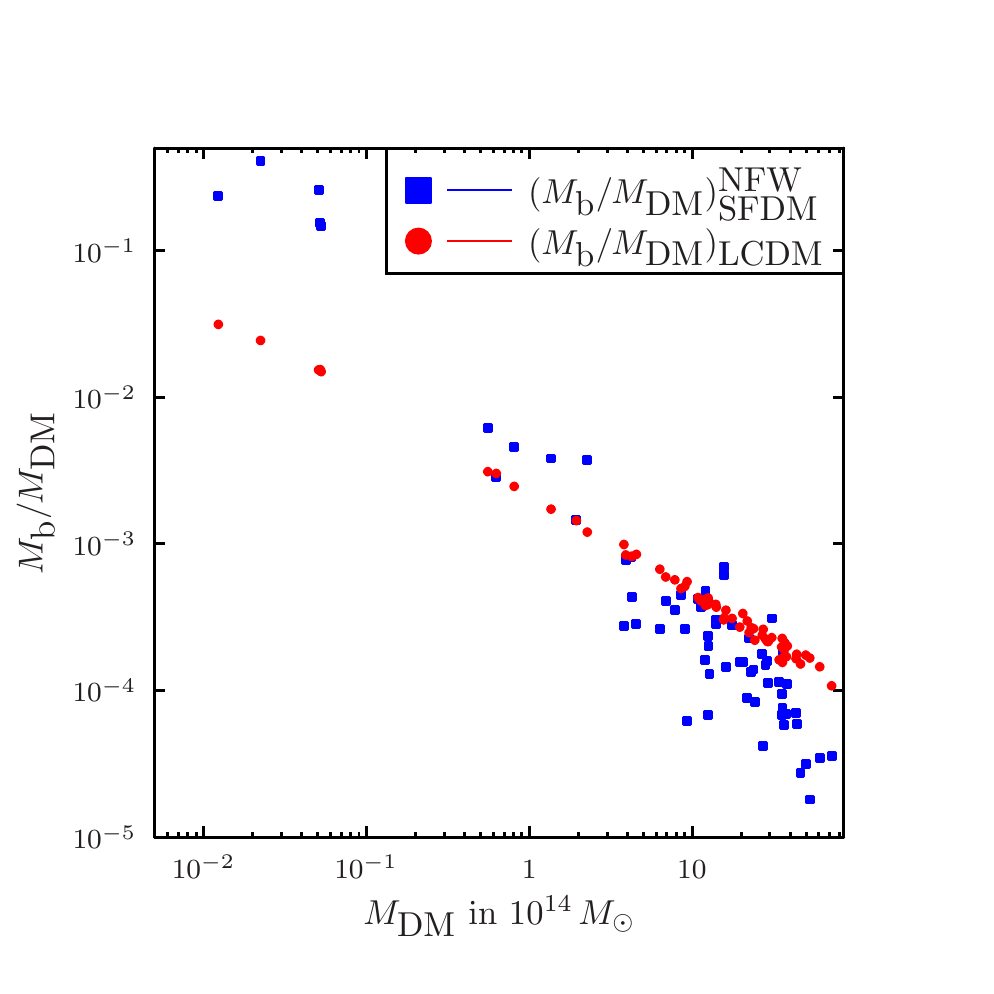}
 \caption{Left: Same as Fig.~\ref{fig:MboverMDM}, but with NFW matching instead of thermal matching and with $ M_\textrm{b} $ and $ \hat{\mu}(r_0) $ fixed at the values from the fitting procedure with thermal matching. Right: Same as Fig.~\ref{fig:MboverMDM}, but with NFW matching instead of thermal matching and with $ \sigma_* $ and $ R_\textrm{E} $ fixed at the values from the fitting procedure with thermal matching.
}
 \label{fig:MboverMDMnfw}
\end{figure}

\subsection{Alternative NFW halo matching}
\label{sec:nfwmatching}

Above, we used the thermal radius $ R_\textrm{T} $ as the radius where we match the superfluid core to an {\sc NFW} tail.
In Ref.~\cite{Berezhiani2017}, an alternative transition radius $ R_\textrm{NFW} $ was proposed. We will now look at this
for completeness to check whether it makes a difference for our conclusions.

This radius $ R_\textrm{NFW} $ is defined by requiring that both the superfluid density and pressure match the respective density and pressure of the {\sc NFW} halo.
Further, Ref.~\cite{Berezhiani2017} did not approximate the {\sc NFW} halo to be proportional to $ 1/r^3 $ but used the full {\sc NFW} profile,
\begin{align}
 \label{eq:nfw}
 \rho_\textrm{NFW}(r) = \frac{\rho_c}{(r/r_s)(1+r/r_s)^2} \,,
\end{align}
where $ \rho_c $ and $ r_s $ are constants.
This matching approach requires three parameters, namely $ \rho_c $, $ r_s$, and $ R_\textrm{NFW} $.
Requiring the density and pressure inside and outside the superfluid phase to match fixes only two of those.

In Ref.~\cite{Berezhiani2017}, the third parameter is fixed by choosing the concentration parameter $ c = r_{200}/r_s $ according to the $\Lambda$CDM Mass-concentration relation of Ref.~\cite{Dutton2014}, with the $ \Lambda $CDM halo mass identified with the calculated $ M_\textrm{DM} $.
In the following, we will call this procedure `{\sc NFW} matching' 
while we refer to the previously discussed procedure as `thermal matching'.

We will now explore how this alternative {\sc NFW} matching affects our results.
In particular, we will compare {\sc NFW}-matching to thermal matching in two different cases.
In the first case, we fix the superfluid core, i.e. the values of $ M_\textrm{b} $ and $ \hat{\mu}(r_0) $, and then match a halo to this superfluid core with both matching procedures.
In the second case, we fix the calculated velocity dispersion and Einstein radius, but allow the superfluid core to vary.

\begin{figure}[t]
 \centering
 \includegraphics[width=.49\textwidth]{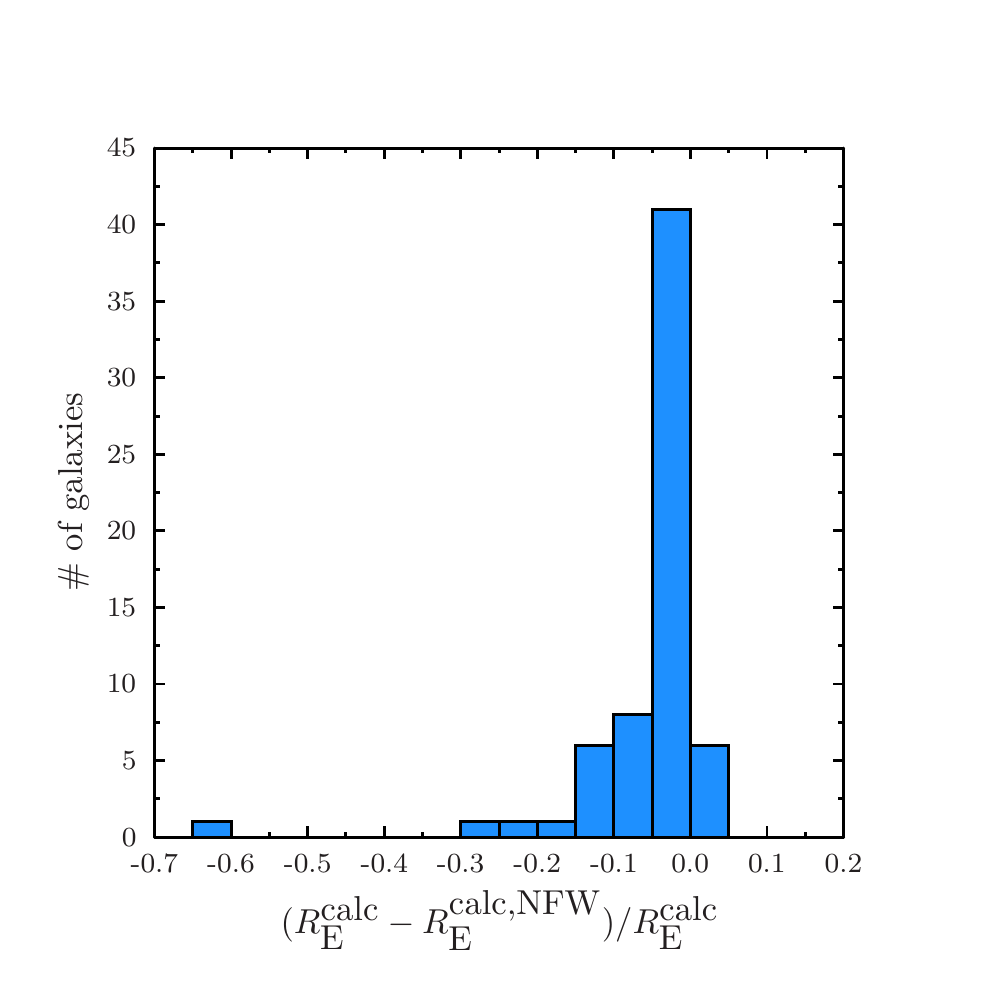}
 \includegraphics[width=.49\textwidth]{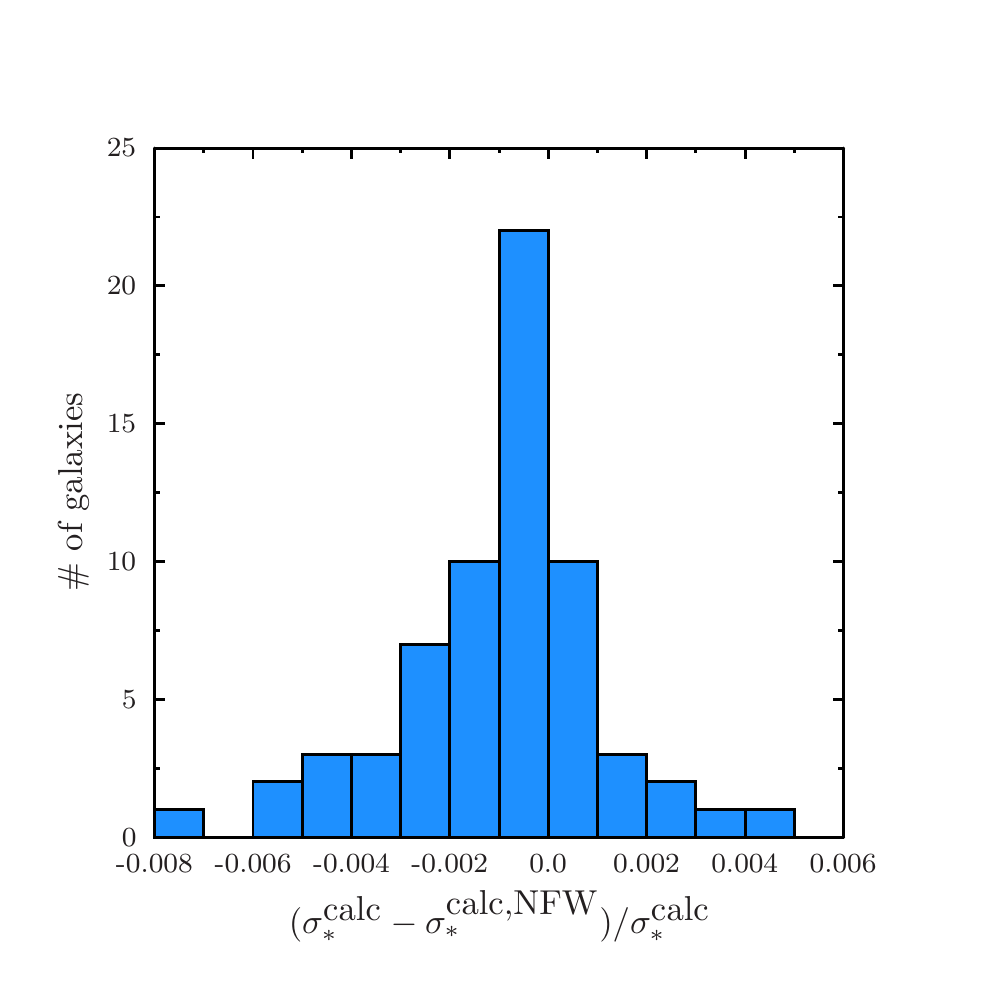}
 \caption{Left: Relative difference in the calculated Einstein radius between the thermal and NFW matching with $ M_\textrm{b} $ and $ \hat{\mu}(r_0) $ fixed at the values from the fitting procedure with thermal matching. Right: Relative difference in the calculated velocity dispersion between the thermal and NFW matching with $ M_\textrm{b} $ and $ \hat{\mu}(r_0) $ fixed at the values from the fitting procedure with thermal matching.}
 \label{fig:histogram-nfw}
\end{figure}

For the case with a fixed superfluid core, we take the values of $ M_\textrm{b} $ and $ \hat{\mu}(r_0) $ as the best-fit values from the thermal matching procedure.
The resulting values of $ M_\textrm{b}/M_\textrm{DM} $ from {\sc NFW} matching are shown in Fig.~\ref{fig:MboverMDMnfw}, left.
Comparing to the result from thermal matching in Fig.~\ref{fig:MboverMDM}, we see that NFW matching gives larger $ M_\textrm{DM} $ values than thermal matching.
Averaging over all galaxies, we find $ M_\textrm{DM}^\textrm{NFW}/M_\textrm{DM}^\textrm{thermal} = 3.4\pm1.3 $, where $ M_\textrm{DM}^\textrm{NFW} $ denotes the values of $ M_\textrm{DM} $ obtained with {\sc NFW} matching and $ M_\textrm{DM}^\textrm{thermal} $ denotes those obtained with thermal matching.
In Fig.~\ref{fig:histogram-nfw}, left, we see that the different matching procedures affect the calculated velocity dispersions only on the sub-percent level.
This is because the velocity dispersions are dominated by the superfluid core which we have kept constant.
In contrast, the Einstein radii receive a larger contribution from the {\sc NFW} halo.
And indeed, the Einstein radii do change above the percent-level when switching to {\sc NFW} matching, see Fig.~\ref{fig:histogram-nfw}, right.
Averaging over all galaxies, the relative difference is 5.2 \% with 9.1 \% standard deviation. 

For the case with fixed calculated velocity dispersion and Einstein radius, we similarly take these fixed values as the best-fit values from the thermal matching procedure.
These fixed values of the velocity dispersion and the Einstein radius can be reached with {\sc NFW} matching by keeping $ M_\textrm{b} $ fixed and adjusting $ \hat{\mu}(r_0) $ in order to match the Einstein radius.
This is possible since the velocity dispersion depends only very weakly on $ \hat{\mu}(r_0) $ as discussed in Sec.~\ref{sec:nonbaryonic}.
Therefore, keeping $ M_\textrm{b} $ fixed keeps the velocity dispersion fixed and adjusting $ \hat{\mu}(r_0) $ effectively adjusts only the Einstein radius.
Indeed, this procedure leads to velocity dispersions which agree with those from thermal matching to at least $ 1\,\% $.
Here, we adjust $ \hat{\mu}(r_0) $ such that the Einstein radii differ by at most $ 0.01\,\textrm{kpc} $.

The resulting values of $ M_\textrm{DM} $ from {\sc NFW} matching are again higher than with thermal matching, but a bit lower than in the case with a fixed superfluid core, see Fig.~\ref{fig:MboverMDMnfw}, right.
We obtain $ M_\textrm{DM}^\textrm{NFW}/M_\textrm{DM}^\textrm{thermal} = 2.8\pm0.8 $.
For all galaxies, the velocity dispersions and Einstein radii obtained with thermal matching can be reproduced with NFW matching.
More precisely, our procedure reproduces the Einstein radii to at least $ 0.01\,\textrm{kpc} $ and the velocity dispersions to at least $ 1\,\% $.
Therefore, {\sc NFW} matching allows to fit the velocity dispersions and Einstein radii at least as good as thermal matching.

To sum up, using {\sc NFW} matching instead of thermal matching has a relatively small effect on the calculated velocity dispersions and Einstein radii.
Further, we can obtain the same velocity dispersions and Einstein radii as with thermal matching by adjusting $ M_\textrm{b} $ and $ \hat{\mu}(r_0) $.
In contrast, {\sc NFW} matching leads to larger values of $ M_\textrm{DM} $ than thermal matching.
However, when interpreting the values of $ M_\textrm{DM} $, it should be kept in mind that $ M_\textrm{DM} $ is quite sensitive to the initial conditions as discussed in Sec.~\ref{sec:mdmsensitivity}.

\subsection{The galaxy J0737+3216}

The galaxy J0737+3216 is the only galaxy in our sample for which we do not obtain a successful fit according to the procedure described in Sec.~\ref{sec:fit}.
The reason we do not obtain a fit is that, given the requirements $ |R_\textrm{E}^\textrm{calc} - R_\textrm{E}^\textrm{meas}| < 0.01\,\textrm{kpc} $ and $ |\sigma_*^\textrm{calc} - \sigma_*^\textrm{meas}| < \sigma_*^\textrm{error} $, no thermalization radius $ R_\textrm{T} $ can be determined since there is no solution to Eq.~\ref{eq:RT}.
This is because $ \Gamma < t_\textrm{dyn}^{-1} $ for all radii which means that there is no superfluid phase in equilibrium.
Therefore, it seems this galaxy should be modeled as having a standard cold dark matter halo at all radii instead of a superfluid halo at small radii and an NFW halo at large radii.

However, note that requiring a superfluid phase and  $ |R_\textrm{E}^\textrm{calc} - R_\textrm{E}^\textrm{meas}| < 0.01\,\textrm{kpc} $, we obtain $ \sigma_*^\textrm{calc} = 314\,\textrm{km/s} $ which is not too far off the measured value $ \sigma_*^\textrm{meas} = 338\pm16\,\textrm{km/s} $.
Indeed, allowing for a $ 5\,\% $ error on $ R_\textrm{E}^\textrm{meas} $, we are able to obtain a successful fit for J0737+3216.
This implies that uncertainties in the measurements or our theoretical modeling could also be the reason for the lack of a successful fit for this galaxy.
Hence, this galaxy does not provide a clear case of a galaxy that should be modeled as not having a superfluid phase in SFDM.

Furthermore we wish to remind the reader that the whole ansatz we are using here for the superfluid in the galaxies relies on
time-independence and spherical symmetry. This may quite possibly simply be a bad ansatz for some galaxies. For this
reason we do not think the one outlier is the interesting part of our analysis. More interesting is that the model
generally seems to work quite well, despite the worries that one might have had about phonons and baryons
reacting to different forces.

\subsection{Discussion}

Strong lensing systems have generally been consistent with the gravitational mass inferred from lensing being equal to the mass inferred from kinematical measurements \cite{Famaey2012}.

In {\sc SFDM}, those two masses differ from each other because the phonon force makes a contribution to the kinematically inferred mass.
Therefore, it is tempting to attribute the success of {\sc SFDM} in fitting Einstein radii and velocity dispersions at the same time to the minor contribution of the phonon force to the velocity dispersion, as discussed in Sec.~\ref{sec:phonon}.

However, as our analysis reveals, the main reason for the success of {\sc SFDM} in fitting the strong lensing systems is a different one: It is the possibility to independently adjust the Einstein radius $ R_\textrm{E} $ and the velocity dispersion $ \sigma_* $ through the two free parameters of the superfluid in each galaxy.
The first of the free parameters in our calculation is the total baryonic mass $ M_\textrm{b} $ which affects both the Einstein radius and the velocity dispersion.
The second parameter $ \hat{\mu}(r_0) $ determines the amount of the non-baryonic gravitational mass.
The important point is that this non-baryonic mass is negligible in calculating the velocity dispersion and only affects the Einstein radius, as discussed in Sec.~\ref{sec:nonbaryonic}.
As a consequence, we can independently adjust the Einstein radius and the velocity dispersion which allows us to fit the Einstein radius and the velocity dispersion at the same time.

It should be added at this point that in {\sc MOND} a similar procedure would not be possible because in this
case one has only one free parameter. 

Finally, there may be other ways of testing the proposal of superfluid dark matter. The ansatz we
studied here must have a UV-completion and as such will almost certainly give rise to new particles.
Such particles can in principle be directly detected if their properties are understood well-enough.
A proposal for a UV-completion has been discussed in \cite{Alexander:2018fjp}, though much
work remains to be done. 

\section{Conclusion}

We have obtained successful fits with reasonable stellar mass-to-light ratios for both the Einstein radius and the velocity dispersion of 64 out of 65 strong
gravitational lenses.
Therefore, we conclude, strong lensing systems -- at least of the type in our sample -- do not seem to pose a challenge for superfluid dark matter.

Furthermore we have demonstrated that the averaged velocity dispersions of the considered galaxies probe the phonon force postulated by {\sc SFDM} mainly at relatively small radii where it is subdominant compared to the gravitational force due to the baryons.
As a result, the phonon force contributes only about $ 10\,\% $ to the calculated velocity dispersion.
For this reason, strong lensing generally -- and not just in our sample -- does not seem to be a promising method to tell apart 
standard cold dark matter from superfluid dark matter.
A cleaner test of the {\sc SFDM} prediction (that the gravitational mass and the mass inferred from kinematics should be different) would have to probe galaxies at smaller baryonic accelerations where the contributions from the phonon force are more important.

\bibliography{sanderslensing}
\bibliographystyle{h-physrev5}

\renewcommand{\arraystretch}{1.3}\setlength{\tabcolsep}{0.35em}\begin{longtable}{l l l l l r@{\hspace{0em}} l l r@{\hspace{0em}} l r}
\caption{Results of our fitting procedure for each galaxy. For $ \hat{\mu}(r_0) $, $ M_\textrm{b} $, $ M_\textrm{b}/L_\textrm{V} $ and $ M_\textrm{DM} $ the notation $ a^{+b}_{-c} $ is used which means that $ a $ is the value which corresponds to the $ \sigma_*^\textrm{calc} $ closest to $\sigma_*^\textrm{meas} $. $ a+b $ and $ a-c $ are the values which correspond, respectively, to the maximum and minimum values which still give a successful fit for $ \sigma_* $ if such values exist, see Sec.~\ref{sec:fit}. $f_\phi \equiv 1-\sigma_*^{\textrm{calc,no}\phi} / \sigma_*^\textrm{calc}$ denotes the fraction the phonon force contributes to $ \sigma_*^\textrm{calc} $.}\\
\label{tab:results}
Lens & $R_\textrm{E}^\textrm{meas}$ & $\sigma_*^\textrm{meas}$ & $\sigma_*^\textrm{calc}$ & $\hat{\mu}(r_0)$ & \multicolumn{2}{l}{$M_\textrm{b}$} & $M_\textrm{b}/L_\textrm{V}$ & \multicolumn{2}{l}{$M_\textrm{DM}$} & $f_\phi$ \\
 & kpc & km/s & km/s & $10^5\,\textrm{eV}$ & \multicolumn{2}{l}{$10^{11}\,M_\odot$} &  & \multicolumn{2}{l}{$10^{14}\,M_\odot$} & \% \\
\hline
J0008-0004 & 6.59 & $193 \pm 36$ & $194_{-35}^{+35}$ & $4.07_{-0.71}^{+0.66}$ & $2.71$ & $_{-1.10}^{+1.40}$ & $2.2_{-0.9}^{+1.1}$ & $9.63$ & $_{-3.53}^{+3.21}$ & $12$ \\
J0029-0055 & 3.48 & $229 \pm 18$ & $229_{-17}^{+6}$ & $1.35_{-0.16}^{+0.54}$ & $3.42$ & $_{-0.55}^{+0.25}$ & $3.6_{-0.6}^{+0.3}$ & $0.41$ & $_{-0.40}^{+1.82}$ & $12$ \\
J0037-0942 & 4.95 & $279 \pm 10$ & $279_{-9}^{+9}$ & $3.77_{-0.30}^{+0.30}$ & $4.27$ & $_{-0.30}^{+0.30}$ & $3.1_{-0.2}^{+0.2}$ & $5.26$ & $_{-1.28}^{+1.36}$ & $9$ \\
J0044+0113 & 1.72 & $266 \pm 13$ & $266_{-11}^{+6}$ & $2.17_{-0.52}^{+1.66}$ & $2.83$ & $_{-0.25}^{+0.15}$ & $3.9_{-0.3}^{+0.2}$ & $1.25$ & $_{-1.24}^{+5.31}$ & $8$ \\
J0157-0056 & 4.89 & $295 \pm 47$ & $285_{-37}^{+0}$ & $1.89_{-0.01}^{+1.08}$ & $8.00$ & $_{-2.35}^{+0.05}$ & $4.6_{-1.4}^{+0.0}$ & $0.03$ & $_{-0.02}^{+4.17}$ & $12$ \\
J0216-0813 & 5.53 & $333 \pm 23$ & $333_{-23}^{+23}$ & $4.63_{-0.90}^{+1.07}$ & $10.85$ & $_{-1.60}^{+1.80}$ & $4.1_{-0.6}^{+0.7}$ & $5.58$ & $_{-3.89}^{+4.68}$ & $10$ \\
J0252+0039 & 4.40 & $164 \pm 12$ & $163_{-10}^{+12}$ & $4.49_{-0.40}^{+0.33}$ & $1.27$ & $_{-0.20}^{+0.25}$ & $1.9_{-0.3}^{+0.4}$ & $11.75$ & $_{-1.69}^{+1.40}$ & $12$ \\
J0330-0020 & 5.45 & $212 \pm 21$ & $212_{-20}^{+19}$ & $3.74_{-0.42}^{+0.42}$ & $2.01$ & $_{-0.40}^{+0.45}$ & $1.8_{-0.3}^{+0.4}$ & $7.52$ & $_{-2.08}^{+2.04}$ & $10$ \\
J0728+3835 & 4.21 & $214 \pm 11$ & $213_{-10}^{+11}$ & $4.42_{-0.48}^{+0.42}$ & $2.41$ & $_{-0.25}^{+0.30}$ & $2.3_{-0.2}^{+0.3}$ & $9.96$ & $_{-1.99}^{+1.77}$ & $10$ \\
J0737+3216 & 4.66 & $338 \pm 16$ & $314$ & $2.37$ & $7.71$ &  & $3.6$ & $0.03$ &  & $10$ \\
J0819+4534 & 2.73 & $225 \pm 15$ & $225_{-15}^{+14}$ & $2.38_{-0.74}^{+1.05}$ & $2.82$ & $_{-0.40}^{+0.40}$ & $4.0_{-0.6}^{+0.6}$ & $3.16$ & $_{-2.41}^{+3.71}$ & $11$ \\
J0822+2652 & 4.45 & $259 \pm 15$ & $259_{-14}^{+15}$ & $3.25_{-0.52}^{+0.59}$ & $4.18$ & $_{-0.50}^{+0.55}$ & $3.5_{-0.4}^{+0.5}$ & $4.51$ & $_{-2.11}^{+2.43}$ & $10$ \\
J0903+4116 & 7.23 & $223 \pm 27$ & $223_{-26}^{+26}$ & $4.29_{-0.50}^{+0.48}$ & $3.93$ & $_{-1.05}^{+1.20}$ & $1.9_{-0.5}^{+0.6}$ & $9.34$ & $_{-2.66}^{+2.55}$ & $12$ \\
J0912+0029 & 4.58 & $326 \pm 12$ & $326_{-12}^{+11}$ & $5.45_{-0.72}^{+0.77}$ & $9.54$ & $_{-0.75}^{+0.75}$ & $5.3_{-0.4}^{+0.4}$ & $9.11$ & $_{-2.90}^{+3.17}$ & $9$ \\
J0935-0003 & 4.26 & $396 \pm 35$ & $370_{-9}^{+0}$ & $3.30_{-0.01}^{+0.35}$ & $13.21$ & $_{-0.80}^{+0.00}$ & $4.1_{-0.2}^{+0.0}$ & $0.02$ & $_{-0.00}^{+1.08}$ & $9$ \\
J0936+0913 & 3.45 & $243 \pm 11$ & $243_{-10}^{+11}$ & $2.22_{-0.42}^{+0.44}$ & $3.45$ & $_{-0.30}^{+0.35}$ & $3.3_{-0.3}^{+0.3}$ & $2.16$ & $_{-1.42}^{+1.59}$ & $11$ \\
J0946+1006 & 4.95 & $263 \pm 21$ & $263_{-20}^{+20}$ & $3.77_{-0.72}^{+0.78}$ & $4.92$ & $_{-0.80}^{+0.90}$ & $5.6_{-0.9}^{+1.0}$ & $6.12$ & $_{-3.03}^{+3.34}$ & $10$ \\
J0956+5100 & 5.05 & $334 \pm 15$ & $334_{-14}^{+15}$ & $3.23_{-0.32}^{+0.51}$ & $8.44$ & $_{-0.80}^{+0.90}$ & $5.6_{-0.5}^{+0.6}$ & $1.34$ & $_{-1.28}^{+2.11}$ & $9$ \\
J0959+4416 & 3.61 & $244 \pm 19$ & $243_{-18}^{+19}$ & $2.78_{-0.89}^{+0.95}$ & $3.81$ & $_{-0.60}^{+0.75}$ & $3.3_{-0.5}^{+0.6}$ & $3.74$ & $_{-3.11}^{+3.60}$ & $11$ \\
J0959+0410 & 2.24 & $197 \pm 13$ & $197_{-12}^{+12}$ & $3.85_{-0.89}^{+1.09}$ & $1.14$ & $_{-0.15}^{+0.15}$ & $4.2_{-0.6}^{+0.6}$ & $8.50$ & $_{-3.21}^{+4.02}$ & $9$ \\
J1016+3859 & 3.13 & $247 \pm 13$ & $248_{-13}^{+11}$ & $3.29_{-0.66}^{+0.84}$ & $2.54$ & $_{-0.30}^{+0.25}$ & $4.0_{-0.5}^{+0.4}$ & $4.96$ & $_{-2.38}^{+3.20}$ & $9$ \\
J1020+1122 & 5.12 & $282 \pm 18$ & $282_{-17}^{+18}$ & $4.45_{-0.65}^{+0.71}$ & $5.06$ & $_{-0.65}^{+0.75}$ & $3.8_{-0.5}^{+0.6}$ & $7.18$ & $_{-2.91}^{+3.11}$ & $9$ \\
J1023+4230 & 4.50 & $242 \pm 15$ & $243_{-15}^{+14}$ & $4.01_{-0.53}^{+0.58}$ & $3.06$ & $_{-0.40}^{+0.40}$ & $3.8_{-0.5}^{+0.5}$ & $7.51$ & $_{-2.23}^{+2.47}$ & $10$ \\
J1029+0420 & 1.92 & $210 \pm 9$ & $210_{-7}^{+7}$ & $1.94_{-0.52}^{+0.63}$ & $1.29$ & $_{-0.10}^{+0.10}$ & $3.5_{-0.3}^{+0.3}$ & $2.31$ & $_{-1.55}^{+2.07}$ & $9$ \\
J1100+5329 & 7.02 & $187 \pm 23$ & $187_{-22}^{+23}$ & $5.73_{-0.48}^{+0.38}$ & $2.55$ & $_{-0.65}^{+0.85}$ & $1.3_{-0.3}^{+0.4}$ & $15.52$ & $_{-2.41}^{+1.91}$ & $11$ \\
J1103+5322 & 2.78 & $196 \pm 12$ & $196_{-12}^{+11}$ & $2.95_{-0.67}^{+0.75}$ & $2.32$ & $_{-0.30}^{+0.30}$ & $3.4_{-0.4}^{+0.4}$ & $5.91$ & $_{-2.36}^{+2.73}$ & $12$ \\
J1106+5228 & 2.17 & $262 \pm 9$ & $262_{-6}^{+3}$ & $1.85_{-0.16}^{+0.41}$ & $1.80$ & $_{-0.10}^{+0.05}$ & $3.4_{-0.2}^{+0.1}$ & $0.52$ & $_{-0.40}^{+1.24}$ & $7$ \\
J1112+0826 & 6.19 & $320 \pm 20$ & $320_{-18}^{+19}$ & $4.31_{-0.40}^{+0.44}$ & $5.75$ & $_{-0.70}^{+0.80}$ & $4.5_{-0.5}^{+0.6}$ & $4.66$ & $_{-2.21}^{+2.35}$ & $8$ \\
J1134+6027 & 2.93 & $239 \pm 11$ & $239_{-10}^{+9}$ & $2.63_{-0.59}^{+0.59}$ & $2.77$ & $_{-0.25}^{+0.25}$ & $4.4_{-0.4}^{+0.4}$ & $3.45$ & $_{-1.97}^{+2.14}$ & $10$ \\
J1142+1001 & 3.52 & $221 \pm 22$ & $222_{-22}^{+20}$ & $4.61_{-1.06}^{+1.32}$ & $2.13$ & $_{-0.45}^{+0.45}$ & $2.4_{-0.5}^{+0.5}$ & $10.13$ & $_{-4.22}^{+5.30}$ & $9$ \\
J1143-0144 & 3.27 & $269 \pm 5$ & $270_{-4}^{+4}$ & $5.05_{-0.26}^{+0.35}$ & $3.36$ & $_{-0.10}^{+0.10}$ & $3.0_{-0.1}^{+0.1}$ & $10.15$ & $_{-1.02}^{+1.33}$ & $8$ \\
J1153+4612 & 3.18 & $226 \pm 15$ & $227_{-14}^{+13}$ & $2.13_{-0.44}^{+0.59}$ & $1.85$ & $_{-0.25}^{+0.25}$ & $3.8_{-0.5}^{+0.5}$ & $2.19$ & $_{-1.54}^{+2.19}$ & $9$ \\
J1204+0358 & 3.68 & $267 \pm 17$ & $266_{-13}^{+17}$ & $3.23_{-0.61}^{+0.52}$ & $2.35$ & $_{-0.25}^{+0.35}$ & $4.5_{-0.5}^{+0.7}$ & $3.92$ & $_{-2.44}^{+2.16}$ & $7$ \\
J1205+4910 & 4.27 & $281 \pm 13$ & $281_{-11}^{+13}$ & $3.50_{-0.55}^{+0.54}$ & $4.60$ & $_{-0.40}^{+0.50}$ & $3.8_{-0.3}^{+0.4}$ & $4.35$ & $_{-2.18}^{+2.18}$ & $9$ \\
J1213+6708 & 3.13 & $292 \pm 11$ & $289_{-7}^{+0}$ & $1.94_{-0.00}^{+0.15}$ & $2.90$ & $_{-0.15}^{+0.00}$ & $3.6_{-0.2}^{+0.0}$ & $0.01$ & $_{-0.00}^{+0.41}$ & $7$ \\
J1218+0830 & 3.47 & $219 \pm 10$ & $219_{-10}^{+9}$ & $4.58_{-0.52}^{+0.61}$ & $2.58$ & $_{-0.25}^{+0.25}$ & $2.8_{-0.3}^{+0.3}$ & $10.47$ & $_{-2.04}^{+2.39}$ & $10$ \\
J1250+0523 & 4.18 & $252 \pm 14$ & $252_{-13}^{+14}$ & $2.37_{-0.36}^{+0.43}$ & $3.05$ & $_{-0.35}^{+0.40}$ & $2.1_{-0.2}^{+0.3}$ & $2.03$ & $_{-1.41}^{+1.73}$ & $10$ \\
J1306+0600 & 3.87 & $237 \pm 17$ & $237_{-16}^{+15}$ & $5.99_{-0.82}^{+0.82}$ & $2.10$ & $_{-0.30}^{+0.30}$ & $3.3_{-0.5}^{+0.5}$ & $14.22$ & $_{-3.42}^{+3.49}$ & $8$ \\
J1313+4615 & 4.25 & $263 \pm 18$ & $263_{-16}^{+17}$ & $4.33_{-0.76}^{+0.79}$ & $3.31$ & $_{-0.45}^{+0.50}$ & $3.9_{-0.5}^{+0.6}$ & $7.66$ & $_{-3.16}^{+3.33}$ & $9$ \\
J1318-0313 & 6.01 & $213 \pm 18$ & $212_{-16}^{+18}$ & $3.98_{-0.44}^{+0.37}$ & $3.29$ & $_{-0.55}^{+0.70}$ & $2.5_{-0.4}^{+0.5}$ & $8.74$ & $_{-2.04}^{+1.74}$ & $12$ \\
J1330-0148 & 1.32 & $185 \pm 9$ & $187_{-8}^{+7}$ & $2.25_{-0.63}^{+1.00}$ & $0.57$ & $_{-0.05}^{+0.05}$ & $5.8_{-0.5}^{+0.5}$ & $3.77$ & $_{-1.99}^{+3.32}$ & $8$ \\
J1402+6321 & 4.53 & $267 \pm 17$ & $267_{-17}^{+16}$ & $4.51_{-0.74}^{+0.84}$ & $4.77$ & $_{-0.65}^{+0.65}$ & $3.6_{-0.5}^{+0.5}$ & $8.31$ & $_{-3.03}^{+3.51}$ & $10$ \\
J1403+0006 & 2.62 & $213 \pm 17$ & $214_{-15}^{+14}$ & $2.94_{-0.85}^{+1.04}$ & $1.66$ & $_{-0.25}^{+0.25}$ & $2.6_{-0.4}^{+0.4}$ & $5.00$ & $_{-2.97}^{+3.83}$ & $10$ \\
J1416+5136 & 6.08 & $240 \pm 25$ & $241_{-23}^{+24}$ & $4.91_{-0.48}^{+0.49}$ & $2.61$ & $_{-0.55}^{+0.60}$ & $2.5_{-0.5}^{+0.6}$ & $10.24$ & $_{-2.64}^{+2.66}$ & $9$ \\
J1420+6019 & 1.26 & $205 \pm 4$ & $203_{-0}^{+4}$ & $1.73_{-0.27}^{+0.00}$ & $1.05$ & $_{-0.00}^{+0.05}$ & $3.2_{-0.0}^{+0.2}$ & $2.05$ & $_{-0.82}^{+0.00}$ & $8$ \\
J1430+4105 & 6.53 & $322 \pm 32$ & $322_{-32}^{+32}$ & $4.43_{-0.81}^{+1.01}$ & $9.63$ & $_{-2.00}^{+2.30}$ & $5.3_{-1.1}^{+1.3}$ & $5.58$ & $_{-4.02}^{+4.85}$ & $10$ \\
J1436-0000 & 4.80 & $224 \pm 17$ & $224_{-15}^{+17}$ & $3.49_{-0.55}^{+0.49}$ & $3.08$ & $_{-0.45}^{+0.55}$ & $2.1_{-0.3}^{+0.4}$ & $6.53$ & $_{-2.32}^{+2.11}$ & $11$ \\
J1443+0304 & 1.93 & $209 \pm 11$ & $208_{-6}^{+11}$ & $2.96_{-0.83}^{+0.49}$ & $0.85$ & $_{-0.05}^{+0.10}$ & $3.0_{-0.2}^{+0.4}$ & $5.15$ & $_{-2.82}^{+1.73}$ & $7$ \\
J1451-0239 & 2.33 & $223 \pm 14$ & $222_{-12}^{+14}$ & $2.28_{-0.84}^{+0.97}$ & $1.67$ & $_{-0.20}^{+0.25}$ & $2.5_{-0.3}^{+0.4}$ & $2.89$ & $_{-2.58}^{+3.35}$ & $9$ \\
J1525+3327 & 6.55 & $264 \pm 26$ & $264_{-26}^{+25}$ & $4.87_{-0.73}^{+0.80}$ & $6.67$ & $_{-1.45}^{+1.55}$ & $2.5_{-0.5}^{+0.6}$ & $9.76$ & $_{-3.48}^{+3.81}$ & $11$ \\
J1531-0105 & 4.71 & $279 \pm 12$ & $278_{-10}^{+12}$ & $3.74_{-0.40}^{+0.37}$ & $3.95$ & $_{-0.30}^{+0.40}$ & $3.1_{-0.2}^{+0.3}$ & $5.28$ & $_{-1.75}^{+1.59}$ & $8$ \\
J1538+5817 & 2.50 & $189 \pm 12$ & $189_{-9}^{+9}$ & $5.20_{-0.69}^{+0.72}$ & $0.95$ & $_{-0.10}^{+0.10}$ & $2.2_{-0.2}^{+0.2}$ & $13.20$ & $_{-2.62}^{+2.77}$ & $9$ \\
J1614+4522 & 2.54 & $182 \pm 13$ & $182_{-13}^{+12}$ & $2.43_{-0.70}^{+0.80}$ & $1.94$ & $_{-0.30}^{+0.30}$ & $3.0_{-0.5}^{+0.5}$ & $4.67$ & $_{-2.35}^{+2.84}$ & $13$ \\
J1621+3931 & 4.97 & $236 \pm 20$ & $237_{-20}^{+18}$ & $5.12_{-0.67}^{+0.77}$ & $3.05$ & $_{-0.55}^{+0.55}$ & $2.2_{-0.4}^{+0.4}$ & $11.32$ & $_{-2.97}^{+3.41}$ & $10$ \\
J1627-0053 & 4.18 & $290 \pm 14$ & $290_{-13}^{+12}$ & $2.38_{-0.27}^{+0.44}$ & $5.19$ & $_{-0.50}^{+0.55}$ & $5.2_{-0.5}^{+0.6}$ & $0.93$ & $_{-0.92}^{+1.65}$ & $9$ \\
J1630+4520 & 6.91 & $276 \pm 16$ & $276_{-15}^{+16}$ & $4.80_{-0.32}^{+0.31}$ & $4.58$ & $_{-0.55}^{+0.60}$ & $3.3_{-0.4}^{+0.4}$ & $8.63$ & $_{-1.78}^{+1.75}$ & $9$ \\
J1636+4707 & 3.96 & $231 \pm 15$ & $231_{-14}^{+15}$ & $3.02_{-0.57}^{+0.64}$ & $2.94$ & $_{-0.40}^{+0.45}$ & $3.1_{-0.4}^{+0.5}$ & $4.83$ & $_{-2.23}^{+2.53}$ & $11$ \\
J1644+2625 & 3.07 & $229 \pm 12$ & $230_{-11}^{+10}$ & $3.77_{-0.68}^{+0.68}$ & $1.94$ & $_{-0.20}^{+0.20}$ & $3.2_{-0.3}^{+0.3}$ & $7.20$ & $_{-2.52}^{+2.62}$ & $9$ \\
J1719+2939 & 3.89 & $286 \pm 15$ & $286_{-14}^{+10}$ & $2.28_{-0.19}^{+0.49}$ & $3.70$ & $_{-0.40}^{+0.30}$ & $5.8_{-0.6}^{+0.5}$ & $0.63$ & $_{-0.62}^{+1.82}$ & $8$ \\
J2238-0754 & 3.08 & $198 \pm 11$ & $198_{-9}^{+11}$ & $5.07_{-0.69}^{+0.57}$ & $1.56$ & $_{-0.15}^{+0.20}$ & $2.3_{-0.2}^{+0.3}$ & $12.68$ & $_{-2.68}^{+2.23}$ & $10$ \\
J2300+0022 & 4.51 & $279 \pm 17$ & $279_{-16}^{+16}$ & $4.83_{-0.77}^{+0.73}$ & $4.22$ & $_{-0.50}^{+0.55}$ & $4.5_{-0.5}^{+0.6}$ & $8.49$ & $_{-3.22}^{+3.16}$ & $9$ \\
J2303+1422 & 4.35 & $255 \pm 16$ & $255_{-16}^{+16}$ & $5.00_{-0.80}^{+0.83}$ & $4.23$ & $_{-0.55}^{+0.60}$ & $3.4_{-0.4}^{+0.5}$ & $10.62$ & $_{-3.27}^{+3.42}$ & $9$ \\
J2321-0939 & 2.47 & $249 \pm 8$ & $250_{-7}^{+7}$ & $3.82_{-0.72}^{+0.63}$ & $3.18$ & $_{-0.20}^{+0.20}$ & $3.7_{-0.2}^{+0.2}$ & $7.16$ & $_{-2.51}^{+2.31}$ & $9$ \\
J2341+0000 & 4.50 & $207 \pm 13$ & $207_{-13}^{+12}$ & $4.80_{-0.48}^{+0.50}$ & $2.59$ & $_{-0.35}^{+0.35}$ & $2.3_{-0.3}^{+0.3}$ & $11.56$ & $_{-2.01}^{+2.13}$ & $11$ \\
J2347-0005 & 6.10 & $404 \pm 59$ & $368_{-23}^{+0}$ & $3.56_{-0.00}^{+0.27}$ & $9.13$ & $_{-1.25}^{+0.00}$ & $4.4_{-0.6}^{+0.0}$ & $0.01$ & $_{-0.00}^{+1.63}$ & $8$ \\
\end{longtable}

\end{document}